\documentclass[aps,preprint]{revtex4-2}
\usepackage{graphicx}
\usepackage{lipsum}
\usepackage{float}
\usepackage{amsmath}
\usepackage{amssymb}
\usepackage{mathtools}
\include{mathmacros}

\begin{document}

\title{
The effect of preferential node deletion on the structure of networks
that evolve via preferential attachment
}

\author{Barak Budnick, Ofer Biham and Eytan Katzav}

\affiliation{
Racah Institute of Physics, 
The Hebrew University, 
Jerusalem 9190401, Israel
}
\begin{abstract}

We present analytical results for the effect of preferential node deletion on
the structure of networks that evolve via node addition and preferential attachment.
To this end, we consider a preferential-attachment-preferential-deletion (PAPD) model,
in which at each time step, with probability $P_{\rm add}$ there is a growth step where
an isolated node is added to the network, followed by the addition of $m$ 
edges, where each edge connects a node selected uniformly at random to a node
selected preferentially in proportion to its degree.
Alternatively, with probability $P_{\rm del}=1-P_{\rm add}$ there is a contraction step,
in which a preferentially selected node is deleted and its links are erased. 
The balance between the growth and contraction processes
is captured by the growth/contraction rate 
$\eta=P_{\rm add}-P_{\rm del}$.
For $0 < \eta \le 1$ the overall process is of network growth,
while for $-1 \le \eta < 0$ the overall process is of network contraction.
Using the master equation and the generating function formalism,
we study the time-dependent degree distribution $P_t(k)$.
It is found that for 
each value of $m>0$ there is a critical value 
$\eta_c(m) = - (m-2)/(m+2)$
such that for
$\eta_c(m) < \eta \le 1$ 
the degree distribution $P_t(k)$ converges towards a stationary distribution
$P_{\rm st}(k)$.
In the special case of pure growth, where $\eta=1$, the model is reduced to
a preferential attachment growth model and $P_{\rm st}(k)$
exhibits a power-law tail, which is a characteristic of scale-free networks.
In contrast, for $\eta_c(m) < \eta < 1$ the distribution $P_{\rm st}(k)$
exhibits an exponential tail, which has a well-defined scale.
This implies a phase transition at $\eta = 1$,
in contrast with the 
preferential-attachment-random-deletion (PARD) model
[B. Budnick, O. Biham and E. Katzav, {\it J. Stat. Mech.} 013401 (2025)],
in which the power-law tail
remains intact as long as $\eta > 0$.
These results illustrate the sensitivity of evolving networks  to
preferential node deletion, in contrast with their robustness to random node deletion.
While for $\eta \ge \max \left\{ \eta_c(m),0 \right\}$ the stationary degree distribution $P_{\rm st}(k)$ lasts indefinitely, 
for $\eta_c(m) < \eta < 0$ (and $m > 2$) it persists for a finite lifetime, until the network vanishes.
It is also found that in the regime of 
$-1 \le \eta \le  \eta_c(m)$ 
the time-dependent degree distribution
$P_t(k)$ does not converge towards a stationary form, but continues to
evolve until the network is reduced to a set of isolated nodes.
These results provide insight on the structure of transient social networks,
such as dating networks and job-seeking platforms,
in which user turnover is intrinsically high. 

\end{abstract}



\maketitle

\newpage

\section{Introduction}

It is well established that the preferential attachment mechanism plays a prominent role
in the growth process of complex networks found in natural, social and technological systems
\cite{Havlin2010,Newman2010,Dorogovtsev2022}.
Previous studies have shown that the preferential attachment process gives rise to networks that
exhibit scale-free structures with power-law degree distributions.
The preferential attachment process is captured by the
Barab\'asi-Albert (BA) model
\cite{Barabasi1999,Albert2002}.
In this model, at each 
time step a new node is added to the network and
forms links to $m$ existing nodes, 
which are selected preferentially, namely with probabilities that are proportional to their degrees
\cite{Barabasi1999,Albert2002}.
The degree distribution of the BA network exhibits a 
power-law tail of the form 
$P(k) \propto k^{- \gamma}$,
where $\gamma=3$ 
\cite{Barabasi1999,Albert2002,Krapivsky2000,Dorogovtsev2000}.
In the case of $\gamma = 3$ the mean degree $\langle K \rangle$ is finite while the 
second moment $\langle K^2 \rangle$ diverges logarithmically.
This is a hallmark of scale-free networks, where a small number of
hubs exhibit disproportionately high degrees compared to other nodes.
A side effect of this property is a great vulnerability to attacks that target
the hubs
\cite{Albert2000,Cohen2001},
in contrast to its resilience to random node failures
\cite{Albert2000,Cohen2000}.

In practice, network growth is often
accompanied by the loss of nodes,
which may be
due to inadvertent failures, deliberate attacks or propagating infections.
Social network contract when the number of users that leave the network is 
larger than the number of users that join 
\cite{Torok2017,Lorincz2019}.
The reasons for leaving the network may be loss of interest, concerns about privacy 
or migration to other social networks.
It is thus important to examine the effect of node deletion on the emerging structure of a growing network.
In this context, several studies have focused on the effect of random node deletion 
on the structure of networks that grow via
preferential attachment  
\cite{Moore2006,Bauke2011,Ghoshal2013}.
It was found that as long as the evolution of the network is dominated by
the growth process (namely, the node addition rate exceeds the deletion rate)
the power-law tail of the degree distribution is maintained.
However, it was recently found that in the opposite regime, in which the rate of node deletion
exceeds the rate of node addition, the degree distribution converges towards a
stationary form which exhibits an exponential tail
\cite{Budnick2025}.
This stationary degree distribution persists for a finite  period of time, until the network vanishes.

The emergence of such intermediate asymptotic state was observed in
an empirical study by Saavedra et al. 
\cite{Saavedra2008},
who examined the business network
of New York City’s garment industry.
In this network nodes represent designers and contractors  
and links denote collaborative productions. 
Over a period of 19 years (1985–2003), 
the network experienced a reduction in size by more than an order of magnitude. 
Despite this decline, the network maintained its structural integrity and resisted 
rapid fragmentation. Its degree distribution remained stationary, displaying a 
truncated power-law behavior. The study revealed that the deletion process 
acted in an anti-preferential manner, rendering low-degree nodes more vulnerable to removal. 
Concurrently, a partial recovery via preferential attachment was observed. 
Such resilience is indicative of the ability of complex networks 
to preserve functionality amid significant contraction.

The case of
preferential node deletion provides a useful framework for the analysis of targeted attacks
\cite{Cohen2001}.
In this process, one selects a random edge and deletes one of its end-nodes together
with its links. This is equivalent to selecting the nodes for deletion in proportion to
their degree, such that the degree of the node to be deleted at time $t$ is sampled from
the distribution 

\begin{equation}
\widetilde P_t(k) = \frac{k}{\langle K \rangle_t} P_t(k), 
\label{eq:Ptilde}
\end{equation}

\noindent
where $P_t(k)$ is the degree
distribution of the network and 

\begin{equation}
\langle K \rangle_t = \sum_{k=1}^{\infty} k P_t(k)
\label{eq:Kmean}
\end{equation}

\noindent
is the mean degree at time $t$.
In fact, preferential node deletion arises naturally in the evolution of complex networks.
For example, consider a random walk (RW) hopping between neighboring nodes in 
a network. The degrees of the nodes visited by the RW are sampled from $\widetilde P_t(k)$,
because the probability that an RW will visit a given node at time $t$ is proportional to its degree
at that time.
This implies that an RW that deletes the nodes visited along its path effectively performs
a process of propagating node deletion where the degrees of the deleted nodes
are selected preferentially
\cite{Tishby2016}.  
Such an RW is in fact a self avoiding (random) walk (SAW).
The preferential selection of nodes is common in
diffusive processes such as epidemic spreading, where the infection of a node may
be viewed as its (preferential) deletion from the subnetwork of susceptible nodes.
Similarly, in social networks, the removal of key influencers can shift opinion formation and
information flow.
This implies that preferential node deletion is an important process in the evolution
of complex networks.

In this paper we study the effect of preferential node deletion on the structure
of networks that evolve via node addition and preferential attachment.
By developing an analytical framework for preferential deletion, we provide insight into how
complex systems respond to targeted disruptions. 
To this end, we consider a preferential-attachment-preferential-deletion (PAPD) model,
in which at each time step, with probability $P_{\rm add}$ 
an isolated node is added to the network, followed by the addition of $m$ edges.
Each one of these edges connects a node selected uniformly at random to a node
selected preferentially in proportion to its degree.
Alternatively, with probability $P_{\rm del}=1-P_{\rm add}$ 
a preferentially selected node is deleted. 
The balance between the growth and contraction processes
is captured by the growth/contraction rate 
$\eta=P_{\rm add}-P_{\rm del}$,
such that for $0 < \eta \le 1$ the overall process is of network growth,
while for $-1 \le \eta < 0$ the overall process is of networks contraction.
Using the master equation and the generating function formalism,
we study the time-dependent degree distribution $P_t(k)$.
We show that for $\eta_c(m) < \eta \le 1$, 
where $\eta_c(m) = -(m-2)/(m+2)$,
the time-dependent degree distribution $P_t(k)$ 
converges towards a stationary distribution
$P_{\rm st}(k)$.
In the special case of pure growth, where $\eta=1$, the model is reduced to
a preferential attachment model and $P_{\rm st}(k)$
exhibits a power-law tail, which is a characteristic of scale-free networks.
In contrast, for $\eta_c(m) < \eta < 1$ the distribution $P_{\rm st}(k)$
exhibits an exponential tail, which has a well-defined scale.
This implies that there is a phase transition at $\eta = 1$.
This is in contrast with the 
preferential-attachment-random-deletion (PARD) model,
in which the power-law tail
remains intact as long as $\eta > 0$
\cite{Budnick2025}.
These results illustrate the  sensitivity of evolving networks to  
preferential node deletion, in contrast to their robustness to random node deletion.
While for $\eta \ge \max \left\{ \eta_c(m),0 \right\}$ the stationary degree distribution $P_{\rm st}(k)$ lasts indefinitely, 
for $\eta_c(m) < \eta < 0$ (and $m > 2$) it persists for a finite lifetime, until the network vanishes.
It is also found that in the regime of $-1 \le \eta \le \eta_c(m)$, the time-dependent degree distribution 
$P_t(k)$ does not converge towards a stationary form but continues to
evolve until the network is reduced to a set of isolated nodes.
This range includes the special case of $\eta=-1$, which corresponds to pure node deletion  
\cite{Tishby2019,Tishby2020}.

The growth step of the PAPD model is motivated by the dynamics of online social networks.
In these networks, the formation of a friendship between two users involves a
two-step process: one user initiates the connection (by sending a friendship request)
and the other user confirms it (by accepting the request).
In spite of this asymmetric process, the resulting connection is symmetric.
The user who initiated the connection is represented by a node selected uniformly at random,
while the user who accepted the friendship request is represented by a node selected preferentially.
This is due to the fact that users who have many friends are more likely to receive additional friendship
requests than users who have few friends.
The model also reflects the fact that in online social networks new friendships are initiated and evolve
gradually over time.

The contraction mechanism of the
PAPD model makes it particularly suitable to describe
transient social networks in which individuals participate with the specific goal of fulfilling a 
time-limited need, after which their engagement with the network typically diminishes or ends. 
Examples include dating networks, where users join to find a romantic partner,
and job-seeking platforms, where users join to secure employment. 
Unlike permanent social networks that emphasize long-term connections, 
transient social networks are goal-oriented in nature. 
Once the objective is met,
such as finding a compatible partner or landing a job,
users typically become inactive or delete their accounts,
leading to naturally high user turnover.
In the context of online social networks, the phenomenon of users
becoming inactive is referred to as ``churn''.
This term originally gained prominence in industries such as telecommunications
and subscription services, where it became a useful metric to measure customer
retention
\cite{Hadden2005,Verbeke2011}.
The preferential attachment reflects the fact that
users in transient social networks tend to prioritize efficiency and success, 
leading them to prefer connections that maximize their chances of quickly 
achieving their goals.
The preferential deletion reflects the fact that highly connected individuals
are more likely to achieve their goals, 
thus eliminating the core need that brought them to the network.
Ironically, high success rates of dating and job-seeking platforms
may lead to high churn rates, posing a challenge for user retention strategies.
This phenomenon is referred to as the ``retention paradox''.

The paper is organized as follows. 
In Sec. II we describe the network evolution model, which
combines growth (via node addition and preferential attachment) and
contraction (via preferential node deletion).
In Sec. III we derive the master equation for the time-dependent 
degree distribution $P_t(k)$. 
In Sec. IV we calculate the stationary generating function $G(u)$ 
and use it to evaluate the stationary 
mean degree $\langle K \rangle$ and the variance ${\rm Var}(K)$. 
In Sec. V we calculate the stationary degree distribution
$P_{\rm st}(k)$ and study the structural phase transition at $\eta=1$.
The results are discussed in Sec. VI 
and summarized in Sec. VII.
In Appendix A we prove a useful property of the probability generating function $G(u)$.
In Appendix B we derive approximations for the mean degree
$\langle K \rangle$
in the limits of 
$\eta \rightarrow \eta_c^{+}(m)$ 
and 
$\eta \rightarrow 1^{-}$. 
In Appendix C we derive an approximation for $\langle K \rangle$
in the special case of $\eta=0$.

\section{The model}  

The model considered in this paper combines growth via
node addition and preferential attachment and contraction
via preferential node deletion.
This model, referred to as the preferential-attachment-preferential-deletion  
(PAPD) model, consists of two processes.
{\it Growth step}: with probability $P_{\rm add}$  
an isolated node (of degree $k=0$) is added to the 
network, followed by the addition of $m$ edges. 
This is done by repeating the following step $m$ times:
each time a node selected uniformly at random is connected 
to a node selected preferentially at random
(namely, with a probability that is proportional to its degree).
{\it Contraction step}: with probability   
$P_{\rm del} = 1 - P_{\rm add}$, 
a preferentially 
selected node is deleted together with its edges.

In this model we also allow non-integer values of $m>0$.
For a non-integer value of $m$, 
we denote by $\lfloor m \rfloor$ the integer part of $m$
(or the floor function).
For non-integer values of $m$, at each time step, with probability
$(\lfloor m \rfloor - m + 1)$ one adds $\lfloor m \rfloor$ new edges,
and with probability $(m - \lfloor m \rfloor)$ one adds $\lfloor m \rfloor + 1$ new edges.
This is the narrowest possible distribution whose random variable takes integer values  
while its mean $m$ has a non-integer value.
Allowing non-integer values of $m$ enables us to present a more complete
phase diagram for the model.

In case that a growth step is selected at time $t$, 
the network size increases from $N_t$
to $N_{t+1}=N_t + 1$.
In each one of the $m$ subsequent edge-addition steps, 
the degrees of the two end-nodes of the new edge increase by $1$.
In some situations it might be impossible to perform a growth step.
This may happen in case that the network includes less than $m$
yet unconnected pairs of nodes in which at least one of the nodes 
in each pair is of degree $k^{\prime} \ge 1$.
In such case the growth step is skipped. 
Such situations typically occur in the contraction dominated case ($\eta<0$)
in the long time limit.
Subsequent contraction steps eventually reduce the network to a set
of isolated nodes.

Alternatively, in case that a contraction step is selected at time $t$, the network size
decreases from $N_t$ to $N_{t+1}=N_t - 1$.
Consider a node of degree $k$, whose $k$ neighbors are of
degrees $k_i^{\prime}$, $i=1,2,\dots,k$.
When such node is selected for deletion,
the degrees of its neighbors are reduced to
$k_i^{\prime} - 1$, $i=1,2,\dots,k$. 
If all the remaining nodes in the network are isolated nodes (whose degree is $k=0$),
the preferential deletion step cannot be performed.
In such case the growth/contraction process is halted.
The network then reaches an absorbing state, in which it consists of
$N_{\rm abs}$ nodes and the degree distribution is 
$P_{\rm abs}(k) = \delta_{k,0}$.

The initial size of the network 
at time $t=0$ is denoted by $N_0$. 
As mentioned above, 
in order to perform the growth/contraction process, 
the initial network must include at least $m$ pairs of nodes
that can be connected according to the rules of the model.
In the case of contraction dominated process ($\eta < 0$),
it is desirable to start from a very large network to allow
sufficient time before the network vanishes.

The expectation value of the network size
at time $t$ is
 
\begin{equation}
N_t = N_0 + \eta t,
\label{eq:Nt}
\end{equation}

\noindent
where
 
\begin{equation}
\eta = P_{\rm add} - P_{\rm del}.
\end{equation}

\noindent
The growth/contraction rate $\eta$ classifies the possible scenarios. 
Pure growth corresponds to $\eta=1$. For $0 <\eta < 1$, the 
overall process is of network growth, while for $-1 \le \eta <0$, 
the overall process is of network contraction. 
In the special case 
of $\eta=0$, the network size remains the same, apart from
possible fluctuations. 
The case of $\eta = -1$ (which corresponds to a pure network contraction process)
was recently studied
\cite{Tishby2019,Tishby2020}.

Expressing the probabilities $P_{\rm add}$ and 
$P_{\rm del}$ in terms of the parameter 
$\eta$, we obtain

\begin{equation}
P_{\rm add} =  \frac{1+\eta}{2} 
\label{eq:Padd}
\end{equation}

\noindent
and

\begin{equation}
P_{\rm del} = \frac{1-\eta}{2}.
\label{eq:Pdel}
\end{equation}

The balance between the growth and contraction processes
can also be expressed by the parameter

\begin{equation}
r = \frac{ P_{\rm add} }{ P_{\rm del} },
\end{equation}

\noindent
which is well defined for $P_{\rm del} > 0$
and takes values in the range $r \ge 0$. 
Expressing $r$ in terms of $\eta$, we obtain

\begin{equation}
r = \frac{1+\eta}{1-\eta},
\label{eq:r}
\end{equation}

\noindent
which is well defined for $\eta < 1$.
Note that the dependence of $r$ on $\eta$ 
is injective, and the inverse relation is given by

\begin{equation}
\eta = \frac{r-1}{r+1},
\end{equation}

\noindent 
such that 
$0 \le r < 1$ 
corresponds to 
$-1 \le \eta < 0$ 
and 
$r \ge 1$ 
corresponds to 
$0 \leq \eta < 1$.

\section{The master equation}

Consider an ensemble of networks, each of size $N_0$ 
at time $t=0$, with an initial degree distribution denoted by $P_0(k)$. 
These networks evolve according to the PAPD model. 
Below, we derive the master equation 
\cite{Vankampen2007,Gardiner2004} that 
describes the time evolution of the  
degree distribution

\begin{equation}
P_t(k) = \frac{N_t(k)}{N_t},
\label{eq:P_t(k)}
\end{equation}

\noindent
where $N_t(k)$, $k=0,1,\dots$, is the number of nodes 
of degree $k$ at time $t$, and 

\begin{equation}
N_t = \sum_{k=0}^{\infty} N_t(k) 
\end{equation}

\noindent
is the size of the network at time $t$.

To derive the master equation, we initially focus on the 
time evolution of $N_t(k)$, expressed in terms of the 
forward difference:

\begin{equation}
\Delta_t N_t(k) = N_{t+1}(k) - N_t(k).
\label{eq:DtNtk}
\end{equation}

\noindent
The addition of an isolated node increases 
the number of nodes of degree $k=0$ by one.
This process is described by the term
$P_{\rm add} \ \delta_{k,0}$, where $\delta_{k,k'}$ is the Kronecker delta.
The effect of the subsequent edge addition on $N_t(k)$ 
consists of two components.
The effect on the nodes selected uniformly at random 
is given by
$m P_{\rm add} [ N_t(k-1) - N_t(k) ]/N_t$,
while the effect on the preferentially selected nodes is given by
$m P_{\rm add} [ (k-1)N_t(k-1) - k  N_t(k) ]/( \langle K \rangle_t N_t)$.

In the case of a preferential deletion step, the primary effect on $N_t(k)$,
due to the deletion of a node of degree $k$ is 
$- P_{\rm del}   kN_t(k) / (\langle K\rangle_t N_t)$.
Below we consider the secondary effect on the neighbors of a deleted node
of degree $k'$, which lose one link each.
The expected number of neighbors of the deleted node, which are of degree $k$ is given by

\begin{equation}
P_{\rm del} \widetilde P_t(k) \sum_{k'=1}^{\infty} k' P_t(k'|k) 
\simeq
P_{\rm del} \sum_{k'=1}^\infty k' \frac{k'N_t(k')}{\langle K\rangle_t N_t}   \frac{k}{\langle K \rangle_t} \frac{N_t(k)}{N_t},
\label{eq:Pkk'Approx}
\end{equation}

\noindent
where $P_t(k'|k)$ is the conditional degree distributions of neighbors of nodes of degree $k$.
In this analysis we neglect the possible effect of degree-degree correlations between pairs
of adjacent nodes, which implies that
$P_t(k'|k) \simeq \widetilde P_t(k')$,
given by Eq. (\ref{eq:Ptilde}). 
This approximation is discussed and numerically justified in Sec. V.
Carrying out the summation over $k'$, it is found that the expected number of nodes
of degree $k$ that lose a link is
$P_{\rm del}\frac{kN_t(k)}{N_t} 
\frac{\langle K^2\rangle_t}{\langle K\rangle^2_t}$,
where

\begin{equation}
\langle K^2 \rangle_t = \sum_{k=1}^{\infty} k^2 P_t(k)
\label{eq:K2}
\end{equation}

\noindent
is the second moment of the degree distribution at time $t$.
Since nodes are discrete entities, the processes of 
node addition and deletion are inherently discrete. 
Consequently, substituting the forward difference 
$\Delta_t N_t(k)$ with a time derivative in the form 
of $dN_t(k)/dt$ entails an approximation. 
The associated error has been demonstrated to 
be of order $1/N_t^2$, rapidly diminishing for 
sufficiently large networks \cite{Tishby2019}. 
Hence, the time evolution of $N_t(k)$ is given by

\begin{eqnarray}
\frac{d}{dt} N_t(k) &=&
P_{\rm add} \left[ \delta_{k,0} + m 
\frac{  N_t(k-1) - N_t(k)  }{N_t} 
+ m \frac{ (k-1) N_t(k-1) - k N_t(k)  }{\langle K \rangle_t N_t}
\right]
\nonumber \\
&+&
P_{\rm del}\left[ -\frac{kN_t(k)}{\langle K\rangle_t N_t} 
+ \frac{\langle K^2 \rangle_t}{\langle K\rangle_t^2} 
\frac{(k+1)N_t(k+1) - kN_t(k) ]}{N_t}\right].
\label{eq:DeltaNtkde0}
\end{eqnarray}

\noindent
To complete the derivation of the master equation we take the time derivative of 
Eq. (\ref{eq:P_t(k)}), which is given by

\begin{equation}
\frac{d}{dt} P_t(k) = \frac{1}{N_t} \frac{d}{dt} N_t(k) - \frac{N_t(k)}{N_t^2} \frac{d}{dt} N_t.
\label{eq:ddtPtk}
\end{equation} 

\noindent
Inserting the time derivative of $N_t(k)$ from Eq. 
(\ref{eq:DeltaNtkde0}) into Eq. (\ref{eq:ddtPtk}) 
and using the fact that $d N_t/dt = \eta$
[which is a conclusion from Eq. (\ref{eq:Nt})],
we obtain the master equation

\begin{eqnarray}
\frac{d}{dt} P_t(k) &=&
\frac{1+\eta}{2N_t} \Biggl\{ \delta_{k,0} + m \left[P_t(k-1) - P_t(k)\right]
+ m \frac{ (k-1) P_t(k-1) - k P_t(k)  }{\langle K \rangle_t} \Biggr\}
\nonumber \\
&+&
\frac{1-\eta}{2N_t}\Biggl\{ -\frac{kP_t(k)}{\langle K\rangle_t N_t} 
+ \frac{\langle K^2 \rangle_t}{\langle K\rangle_t^2} 
\left[ (k+1)P_t(k+1) - kP_t(k)\right]\Biggr\}-\eta \frac{P_t(k)}{N_t},
\label{eq:dP(t)/dtRC0}
\end{eqnarray}

\noindent
where we express $P_{\rm add}$ and $P_{\rm del}$
in terms of $\eta$, using Eqs. (\ref{eq:Padd}) and (\ref{eq:Pdel}).
In Eq. (\ref{eq:dP(t)/dtRC0}), the network size $N_t$ is given by
Eq. (\ref{eq:Nt}),
the mean degree $\langle K \rangle_t$ is given by Eq. (\ref{eq:Kmean})
and the second moment $\langle K^2 \rangle_t$ is given by 
Eq. (\ref{eq:K2}).
The rate coefficients on the right-hand side of Eq. (\ref{eq:dP(t)/dtRC0})
are time-dependent via $N_t$, $\langle K \rangle_t$ and $\langle K^2 \rangle_t$.
While the network size $N_t$ depends explicitly on time, the moments $\langle K \rangle_t$
and $\langle K^2 \rangle_t$ depend implicitly on time via the degree distribution $P_t(k)$.
This renders the master equation nonlinear, unlike ordinary master equations,
which exhibit constant coefficients and are linear in $P_t(k)$.

The master equation 
(\ref{eq:dP(t)/dtRC0})
consists of a set of coupled ordinary differential equations for 
the degree distribution $P_t(k)$, $k=0,1,2,\dots$
\cite{Vankampen2007,Gardiner2004}.
It accounts for the time evolution of $P_t(k)$
over an ensemble of networks,
which start at time $t=0$ from initial networks of size $N_0$  
with degree distribution $P_0(k)$ and are exposed to 
the same growth and contraction processes.
The master equation is commonly solved by direct numerical integration. 
Since the rate coefficients of Eq. (\ref{eq:dP(t)/dtRC0})
are time-dependent, they must be updated as time evolves.
However, in the large network limit they can be considered as slow variables and may thus
be updated once every several time steps.

In the regime of overall network growth, where $\eta>0$,
in the long-time limit, $P_t(k)$ is expected to converge towards
a stationary degree distribution $P_{\rm st}(k)$, which remains
fixed as the network continues to grow.
Surprisingly, it was found that in related models, 
such as the PARD model
\cite{Budnick2025}
and the random-attachment-random-deletion (RARD) model
\cite{Budnick2022},
$P_t(k)$ converges
towards a stationary degree distribution even in the regime of
overall network contraction, where $-1 < \eta < 0$.
In this case, the stationary phase exhibits a finite lifetime
and is diminished when the network vanishes.
In this scenario, $P_{\rm st}(k)$ 
is considered an intermediate asymptotic state 
\cite{Barenblatt2003, Barenblatt1996}. 
Intermediate asymptotic states emerge at time scales that are 
long enough for such structures to develop but shorter than 
the time scales at which the entire system disintegrates. 
Increasing the initial size of the system can extend these 
intermediate time scales arbitrarily, justifying the term 'asymptotic'. 
Since the degree distribution is a local structural property of the network, 
this implies that even when the network shrinks, the local neighborhood 
of a typical node remains invariant at intermediate time scales. 

In the analysis below we use analytical methods to calculate
the stationary degree distribution $P_{\rm st}(k)$.
Since the analytical methods involve approximations, we verify
the validity of the analytical results by comparing them with
the results obtained from computer simulations.
In the simulations presented below, the initial network 
is an Erd{\H o}s-R\'enyi (ER) network
\cite{Erdos1959,Erdos1960,Erdos1961}
of size $N_0$, with mean degree $c = \langle K \rangle_0$.
ER networks are
random networks in which
each pair of nodes 
is connected
with probability
$p=c/(N_0-1)$.
The ER network ensemble is a maximum entropy ensemble under the condition
that the mean degree 
$c$  
is constrained.
The degree distribution of an ER network follows a
Poisson distribution of the form
\cite{Bollobas2001}
 
\begin{equation}
P(k) = \frac{e^{-c} c^k}{k!}.
\label{eq:poisson}
\end{equation}

\section{The mean degree and the variance}

To calculate the mean degree 
$\langle K \rangle$ and variance ${\rm Var}(K)$ 
of the stationary degree distribution, 
we use the generating function 

\begin{equation}
G(u) = \sum_{k=0}^{\infty} u^k P_{\rm st}(k),
\label{eq:GuDef}
\end{equation}

\noindent
which is the Z-transform of $P_{\rm st}(k)$
\cite{Phillips2015}.
The generating function satisfies $G(1)=1$ 
and $0 \le G(u) \le 1$ for $u \in [0,1]$.
Setting $dP_t(k)/dt=0$ in 
Eq. (\ref{eq:dP(t)/dtRC0}) 
we obtain a set of coupled algebraic equations for $P_{\rm st}(k)$,
in which the time-dependent moments $\langle K \rangle_t$
and $\langle K^2 \rangle_t$ are replaced by the stationary moments
$\langle K \rangle$ and $\langle K^2 \rangle$, respectively.
Multiplying these equations
by $u^k$ and summing up over $k$, we obtain an ordinary 
differential equation for $G(u)$,
which takes the form

\begin{equation}
\left[ \frac{rm}{\langle K\rangle}u(u-1) 
- \frac{\langle K^2 \rangle}{\langle K\rangle^2}(u-1) 
- \frac{u}{\langle K\rangle} \right]\frac{ d G(u)}{d u} 
+ \left[ rm(u-1)-r+1 \right]G(u)+r = 0.
\label{eq:diffeq0a}
\end{equation}

\noindent
Note that Eq. (\ref{eq:diffeq0a}) is nonlinear, because
both $\langle K \rangle$ and $\langle K^2 \rangle$
depend on the stationary degree distribution $P_{\rm st}(k)$.
In fact, 

\begin{equation}
\langle K \rangle = \frac{ d G(u) }{ du } \bigg\vert_{u=1},
\end{equation}

\noindent
and

\begin{equation}
\langle K^2 \rangle = 
\frac{ d^2 G(u) }{ du^2 } \bigg\vert_{u=1} 
+ 
\frac{ d G(u) }{ du } \bigg\vert_{u=1}.
\end{equation}

\noindent
This renders the equation non-linear.

Differentiating Eq. (\ref{eq:diffeq0a}) with respect to $u$ and
plugging $u=1$, 
we obtain

\begin{equation}
\frac{\langle K^2\rangle}{\langle K\rangle^2} 
= 
\frac{rm}{\langle K\rangle}-\frac{r-1}{2}.
\label{eq:KK^2}
\end{equation}

\noindent
Inserting  
${\langle K^2 \rangle}/{\langle K \rangle^2}$
from 
Eq. (\ref{eq:KK^2}) 
into Eq. (\ref{eq:diffeq0a}) 
and rearranging terms,
we obtain

\begin{equation}
\left[
\frac{ r m }{\langle K \rangle} u^2
- \left( \frac{2 r m + 1}{\langle K \rangle} - \frac{r - 1}{2} \right) u
+ \frac{ r m }{\langle K \rangle} - \frac{ r - 1 }{2}
\right]
\frac{ d G(u)}{d u} 
+ \left[ rm(u-1)-r+1 \right]G(u)+r = 0.
\label{eq:diffeq0b}
\end{equation}

\noindent
The roots of the pre-factor of $d G(u)/du$ 
in Eq. (\ref{eq:diffeq0b})
are given by

\begin{equation}
u_{\pm} = \frac{2 - \langle K \rangle (r-1)
+
4 rm \pm \sqrt{\left[2 - \langle K \rangle (r-1) \right]^2 + 16 r m}}{4 r m}.
\label{eq:upum}
\end{equation}

\noindent
These are the two regular-singular points
\cite{Bender1999}
of the differential equation
(\ref{eq:diffeq0b}).
Rearranging terms in Eq. (\ref{eq:diffeq0b}), we obtain

\begin{equation}
\frac{ d G(u)}{d u} 
= 
\langle K \rangle \frac{ u - \left( 1 + \frac{r-1}{rm} \right)}{ (u_{+} - u) (u - u_{-}) } G(u) 
+ 
\frac{\langle K \rangle}{m} \frac{1}{ (u_{+} - u)(u - u_{-}) }.
\label{eq:diffGu}
\end{equation}

Using the fact that the parameters $r$ and $m$ are poitive and the mean degree 
satisfies $0 < \langle K\rangle < m$, 
one observes that the roots $u_{+}$ and $u_{-}$ satisfy
$u_{+} > 1$ and $0 < u_{-} < 1$.
Below we
solve Eq. (\ref{eq:diffGu}), with the aim of expressing
the generating function $G(u)$ in terms of the 
stationary mean degree $\langle K \rangle$.
To this end, we express the solution of 
Eq. (\ref{eq:diffGu}) 
in the form

\begin{equation}
G(u) = G_h(u) + G_p(u),
\label{eq:Gsum}
\end{equation}

\noindent
where $G_h(u)$ is a homogeneous solution and $G_p(u)$ is a
particular solution.

To find a particular solution $G_p(u)$, we use
the integration factor

\begin{equation}
\mu(u) = (u_+-u)^{\alpha_+}(u-u_-)^{\alpha_-},
\label{eq:Muu}
\end{equation}

\noindent
where

\begin{equation}
\alpha_\pm = \frac{\langle K\rangle}{2} 
\left[ 1 \pm \frac{6-4r-\langle K\rangle(r-1)}{2rm(u_+-u_-)} \right].
\label{eq:Alphapm}
\end{equation}

\noindent
Note that the second term in the square parentheses on the
right hand side of Eq. (\ref{eq:Alphapm}) 
may be either positive or negative for different values of $r$ and $m$.
As a result, for some values of these parameters $\alpha_{+}$ may be
larger than $\alpha_{-}$, while for other values 
$\alpha_{-}$ may be larger than $\alpha_{+}$.
Multiplying both sides of Eq. (\ref{eq:diffGu}) by 
$\mu(u)$ and rearranging terms,
we obtain

\begin{equation}
\frac{d}{du}\left[G_p(u)\mu(u)\right] 
= \frac{\langle K\rangle}{m}
(u_+-u)^{\alpha_{+} - 1} (u-u_-)^{\alpha_{-} - 1}.
\label{eq:diffGpu}
\end{equation}

In Appendix A we prove that 
the exponent $\alpha_-$  
must satisfy
$\alpha_- > 0$
for any choice of parameters for which the stationary
degree distribution $P_{\rm st}(k)$ exists. 
This property enables us to integrate both sides of 
Eq. (\ref{eq:diffGpu}) 
in the interval $[u_-,u]$ 
for any value of $u_- < u < u_+$,
without crossing any singularity. 
Using the fact that $\mu(u_-)=0$ we obtain

\begin{equation}
G_p(u) = \frac{\langle K\rangle}{m} 
\frac{ 1 }{ (u_+-u)^{ \alpha_+} (u-u_-)^{ \alpha_-} }
\int_{u_-}^u 
d x   (u_{+} - x)^{ \alpha_{+} - 1} (x-u_-)^{ \alpha_{-} - 1} .
\label{eq:GpuInt}
\end{equation}

\noindent
We now consider the homogeneous equation, which takes the form

\begin{equation}
\frac{d}{du}\left[G_h(u)\mu(u)\right] 
= 0.
\label{eq:diffGhu}
\end{equation}

\noindent
The solution of the homogeneous equation is given by

\begin{equation}
G_h(u) =  \frac{ A }{ (u_+-u)^{ \alpha_+} (u-u_-)^{  \alpha_-} } ,  
\label{eq:Ghu}
\end{equation}

\noindent
where $A$ is an integration constant.
Since $\alpha_- > 0$, the homogeneous part of the generating fuction $G_h(u)$,
given by Eq. (\ref{eq:Ghu}), diverges in the limit of $u \rightarrow u_-$.
This contradicts the condition that $G(u)$ must be bounded in the
interval $0 \le u \le 1$.
This implies that the integration constant must satisfy $A=0$.
Hence, the generating function $G(u)$ is simply the particular generating function $G_p(u)$. 
Replacing the integration variable $x$ by $y=(x-u_-)/(u-u_-)$ and using the integral 
representation of the hypergeometric function 
(equation 15.6.1 in Ref. \cite{Olver2010}),  
Eq. (\ref{eq:GpuInt}) 
can be written in the form

\begin{equation}
G(u) = \frac{\langle K\rangle}{m\alpha_-(u_+-u_-)}\left( \frac{u_+-u_-}{u_+-u} \right)^{\alpha_+}
\, _2F_1 \left[ \left.
\begin{array}{c}
1-\alpha_+, \alpha_- \\
\alpha_-+1
\end{array}
\right|    \frac{u-u_-}{u_+-u_-}
\right],
\label{eq:GuHyper}
\end{equation}

\noindent
where

\begin{equation}
_2F_1 
\left[ \left.
\begin{array}{c}
a, b \\  
c
\end{array}
\right| z 
\right] =
\sum_{n=0}^{\infty} 
\frac{ (a)_n (b)_n }{ (c)_n } \frac{ z^n }{ n! },
\label{eq:2F1}
\end{equation}

\noindent 
is the hypergeometric function and

\begin{equation}
(x)_n = 
\left\{
\begin{array}{ll}
1 & n=0 \\
\prod\limits_{i=0}^{n-1} (x+i) & n \ge 1
\end{array}
\right.
\end{equation}

\noindent
is the rising Pochhammer symbol
\cite{Olver2010}.

Plugging $u=1$ into Eq. (\ref{eq:GuHyper}) and relying on the fact that
$G(1)=1$, we obtain an implicit equation for $\langle K\rangle$,
which takes the form

\begin{equation}
\frac{\langle K\rangle}{m\alpha_-(u_+-u_-)}\left( \frac{u_+-u_-}{u_+-1} \right)^{\alpha_+}
\, _2F_1 \left[ \left.
\begin{array}{c}
1-\alpha_+, \alpha_- \\
\alpha_-+1
\end{array}
\right|    \frac{1-u_-}{u_+-u_-}
\right]=1.
\label{eq:K_SCE}
\end{equation}

\noindent
Solving Eq. (\ref{eq:K_SCE}) provides the mean degree 
$\langle K \rangle$ for 
any desired choice of the parameters $r>0$ and $m > 0$. 
It turns out that for some values of $r, m > 0$ the value of the mean degree
$\langle K \rangle$, obtained from Eq. (\ref{eq:K_SCE}), is positive,
while for other values it is negative.

Clearly, the stationary solution exists only in the regime in which 
the mean degree satisfies
$\langle K \rangle > 0$.
Below we calculate, for a given value of $m$, 
the lowest value of the parameter $r$
for which a stationary solution exists.
To this end we set $\langle K \rangle=0$  
and use 
Eq. (\ref{eq:K_SCE}) 
to obtain the 
critical value
$r_c(m)$.
Plugging $r=r_c(m)$ 
and $\langle K \rangle = 0$ into 
$u_\pm$ [Eq. (\ref{eq:upum})] and 
$\alpha_\pm$ [Eq. (\ref{eq:Alphapm})] 
and inserting them into 
Eq. (\ref{eq:K_SCE}) 
we find that $r_c(m)$ satisfies

\begin{equation}
\frac{ 2 r_c(m) }{ 2 r_c(m) - 3 + \sqrt{ 4 m r_c(m) + 1 } } = 1,
\label{eq:rcEq}
\end{equation}

\noindent 
which yields

\begin{equation}
r_c(m) = \frac{2}{m}.
\label{eq:rc}
\end{equation}

\noindent 
In terms of the growth/contraction rate $\eta$, 
the critical value $\eta_c(m)$ is given by

\begin{equation}
\eta_c(m) = - \frac{m-2}{m+2}.
\label{eq:eta_c}
\end{equation}

\noindent
Inverting Eq. (\ref{eq:eta_c}), one can find, for each value of the growth/contraction rate $\eta$,
the corresponding value of the parameter $m$, for which the boundary between the stationary
and disintegrated states is at the given value of $\eta$.
It is found that

\begin{equation}
m_c(\eta) = 2 \frac{1-\eta}{1+\eta}.
\label{eq:m_c_eta}
\end{equation}

\noindent
For a given value of $m$, when 
$\eta$ is increased the growth rate of the network is enhanced.
This implies that
$d \langle K \rangle/d \eta  > 0$.
Therefore, the mean degree $\langle K \rangle$,
obtained from Eq. (\ref{eq:K_SCE}),
is positive for $\eta>\eta_c$,
while for $\eta < \eta_c$ it is negative, which is non-physical.

This result implies that in the regime of $\eta < \eta_c$ there is no
stationary solution $P_{\rm st}(k)$ for the degree distribution.
It means that for $\eta < \eta_c$ the time dependent degree distribution $P_t(k)$
keeps evolving until the whole network disintegrates
and $\langle K \rangle_t \rightarrow 0$.
In other words, there is no solution  
$G(u)$ of Eq. (\ref{eq:diffeq0a}) which satisfies the conditions $G(1)=1$ (normalization)
and 
$dG(u)/du \vert_{u=1} > 0$ 
(positive mean degree). 
We thus conclude that the boundary line $\eta = \eta_c(m)$ is a line of critical points.

When $\eta < \max \left\{ \eta_c(m), 0 \right\}$, 
there exists a time $t_{\rm dis}$,
referred to as the disintegration time,
at which the network reaches 
a state where all the remaining nodes 
are isolated nodes of degree $k=0$. 
Since both the edge-addition mechanism and the node deletion mechanism 
require nodes with degree $k>0$, the process is forced to 
stop and the network remains in its disintegrated phase.
This final state is referred to as an absorbing state
\cite{Kemeny1976}.
From Eq. (\ref{eq:Nt}) one can easily see that for $\eta < 0$ the
network disintegrates after a finite time, which  
satisfies $t_{\rm dis} < N_0 / | \eta |$.

The special case of $\eta = -1$ (pure node deletion) was studied in Refs. 
\cite{Tishby2019,Tishby2020}.
It was shown that in such case, as the network size
decreases, the degree distribution 
converges towards a time-dependent Poisson distribution with a monotonically
decreasing mean degree. 
After completion of the contraction process the network is reduced to
a set of isolated nodes, which is an absorbing state.


In Appendix B we derive approximated expressions for $\langle K\rangle$ 
in the limits of 
$\eta\rightarrow\eta_c^{+}(m)$ 
and 
$\eta\rightarrow 1^-$. 
More specifically, we find that in the limit of 
$\eta\rightarrow\eta_c^{+}(m)$ 
the mean degree is given by

\begin{equation}
\langle K\rangle = 
\frac{(m+2)^2}{m-2+12\ln\left(3/2\right)} [\eta-\eta_c(m)] 
+ \mathcal{O} [\eta-\eta_c(m)]^2,
\label{eq:K_eta_c}
\end{equation}

\noindent 
while in the limit of
$\eta\rightarrow 1^-$
it is given by

\begin{equation}
\langle K\rangle = 2m - \left\{ 2(2m+1)
\left[\ln\left( \frac{2m}{1-\eta} \right) - H_{2m+2} \right]
-7m-\frac{1}{m+1} \right\}\left(1-\eta\right) 
+ \mathcal{O}(1-\eta)^2,
\label{eq:K_1}
\end{equation}

\noindent 
where $H_{n}$ is the $n$th harmonic number
\cite{Olver2010}.

In Ref. \cite{Budnick2025} there is a closed-form expression for $P_{\rm st}(k)$ in the special case of 
pure growth ($\eta=1$). It is also shown that in this case $\langle K\rangle = 2m$. 
Hence, Eq. (\ref{eq:K_1}) implies that applying an infinitesimal rate of 
preferential node deletion does not affect $\langle K\rangle$ in a 
discontinuous manner.

In the special case of $\eta=0$ the expectation value of the network size remains fixed.
In this case, it is possible to obtain a closed-form approximation for the mean degree
$\langle K \rangle$ as a function of $m$, as shown in Appendix C
[Eq. (\ref{eq:Kmean_for_eta0})].

In Fig. \ref{fig:1} we present analytical results (solid lines) 
for the mean degree $\langle K\rangle$ 
of the stationary degree distribution $P_{\rm st}(k)$, 
obtained from numerical solution of Eq. (\ref{eq:K_SCE})
for $m=6$, $10$ and $15$ (bottom to top).
Note that at $\eta=1$ the mean degree satisfies
$\langle K \rangle = 2m$.
To validate our calculations, simulation results (symbols) are also 
shown, and are found to be in very good agreement with the 
corresponding analytical results. We also present the 
asymptotic approximations for $\langle K\rangle$ 
for the case $m=15$ around the two critical points: 
$\eta=\eta_c(m)$ (black dashed line), obtained from 
Eq. (\ref{eq:K_eta_c}), and $\eta=1$ (dotted line), 
obtained from Eq. (\ref{eq:K_1}). 
It appears that the 
validity of both approximations is confined only to a 
narrow vicinity around the critical points, rendering 
them impractical for evaluating $\langle K\rangle$ 
for most values of $\eta$.
However, they reveal information about the scaling behavior 
of $P_{\rm st}(k)$, which will be useful in the analysis below.

\begin{figure}
\begin{center}
\includegraphics[width=7cm]{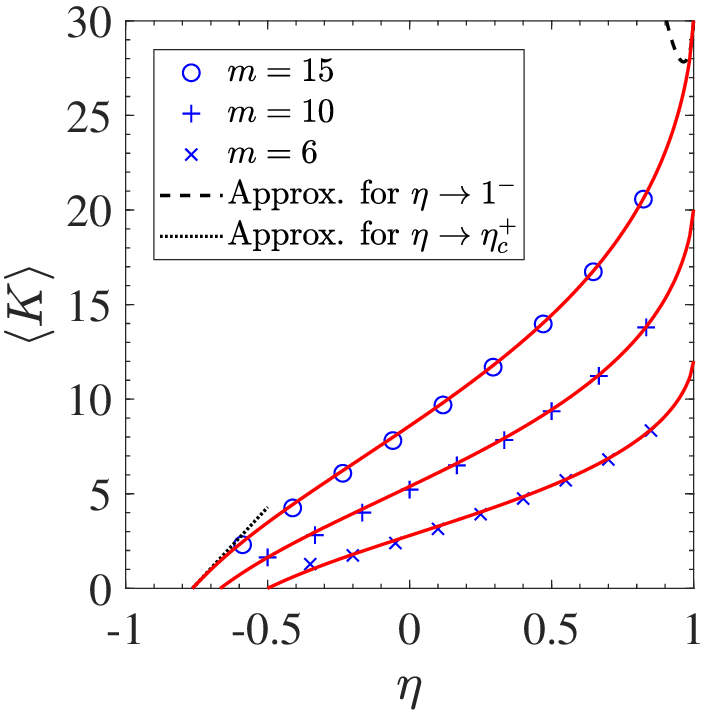}
\hspace{0.5 cm}
\end{center}
\caption{
(Color online)
Analytical results (solid lines) for the mean degree 
$\langle K\rangle$ of the stationary degree distribution $P_{\rm st}(k)$,
obtained from a numerical solution of Eq. (\ref{eq:K_SCE}),
as a function of the growth/contraction rate $\eta$. 
The results are presented for $m=6$, $10$ and $15$ (bottom to top). 
For each value of $m$, a  (positive) solution exists for $\eta_c(m) < \eta \leq 1$, 
where $\eta_c(m)$ is given by Eq. (\ref{eq:eta_c}).
The simulation results (symbols) are in very good agreement 
with the corresponding analytical results. 
For $0 < \eta < 1$ the initial network is an ER network 
that consists of $N_0=200$ nodes 
with mean degree $\langle K \rangle_0 = 3$. 
The results are shown at time $t=(N_t - N_0)/\eta$,
at which the network size has
reached $N_t=20,000$ nodes.
For $\eta=0$ the initial network is an ER network that 
consists of $N_0=20,000$ nodes with mean degree
$\langle K \rangle_0 = 8$. 
The results are shown at time $t=80,000$. 
For $\eta_c < \eta < 0$ the initial network is an ER network that consists 
of $N_0=20,000$ nodes with mean degree
$\langle K \rangle_0 = 8$. 
The results are shown at time $t = (N_0 - N_t)/| \eta |$,
at which the network size declined to
$N_t=2,000$ nodes. 
We also present the results of a series expansion to leading order in $\eta$
for $\langle K \rangle$ vs. $\eta$, 
in the case of $m=15$, in the vicinity of $\eta_c(m=15)$ 
(dashed line) and $\eta=1$ (dotted line), obtained from
Eqs. (\ref{eq:K_eta_c}) and (\ref{eq:K_1}), respectively. 
These results are found to be valid only in a narrow regime
near $\eta=\eta_c(m=15)$ and $\eta=1$.
}
\label{fig:1}
\end{figure}

To obtain the variance 
${\rm Var}(K) = \langle K^2 \rangle - \langle K \rangle^2$
of the stationary degree distribution $P_{\rm st}(k)$,
we use Eq. (\ref{eq:KK^2}) to express 
$\langle K^2\rangle$ 
in terms of $\langle K\rangle$. 
We obtain

\begin{equation}
{\rm Var}(K) = \frac{\langle K\rangle}{1-\eta}
\left[ m\left(1+\eta\right)-\langle K\rangle \right].
\label{eq:VarK}
\end{equation}

\noindent 
It is worth noting that, for the variance to remain 
non-negative, the mean degree must satisfy 
$\langle K\rangle < m (1+\eta)$. 
This requirement is clearly satisfied from 
the following considerations.
If we were to replace 
the preferential node deletion with a random node 
deletion mechanism, 
the mean degree would be given by 
$\langle K \rangle_{\rm PARD} = m ( 1+\eta )$ 
\cite{Moore2006,Bauke2011,Ghoshal2013,Budnick2025}. 
Since the contraction mechanism of the PAPD model
targets high-degree nodes 
with a higher probability, 
on average, each contraction step leads to the deletion
of more edges compared to the PARD model
with the same values of $m$ and $\eta$. 
Consequently, 
$\langle K\rangle < \langle K\rangle_{\rm PARD}$,
which implies that the expression for the variance 
on the right hand side of
Eq. (\ref{eq:VarK}) is positive.

\begin{figure}
\begin{center}
\includegraphics[width=7cm]{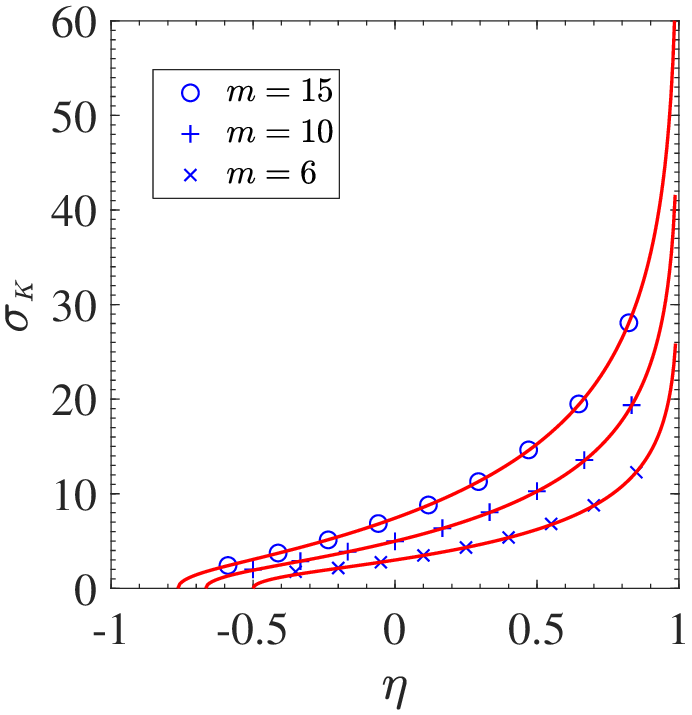}
\hspace{0.5 cm}
\end{center}
\caption{
(Color online)
Analytical results (solid lines) for the standard deviation 
$\sigma_K = \sqrt{ {\rm Var}(K) }$ of the stationary degree distribution $P_{\rm st}(k)$,  
as a function of $\eta$, 
where ${\rm Var}(K)$ is obtained
from Eq. (\ref{eq:VarK}). 
The results are presented for $m=6$, $10$ and $15$
(bottom to top). 
The simulation results (symbols) are in very good agreement with the 
corresponding analytical results. 
The parameters of the simulations are the same as in Fig. \ref{fig:1}.
}
\label{fig:2}
\end{figure}

In Fig. \ref{fig:2} we present 
analytical results (solid lines) for the standard deviation 
$\sigma_K = \sqrt{ {\rm Var}(K)}$ of the stationary degree distribution
$P_{\rm st}(k)$,  
as a function of $\eta$,
obtained from Eq. (\ref{eq:VarK}).
Results are presented for $m=6$, $10$ 
and $15$ (bottom to top). 
The analytical results are in very good agreement with the 
results obtained from computer simulations (symbols).

\section{The stationary degree distribution $P_{\rm st}(k)$}

To obtain a closed-form expression for the stationary degree distribution
$P_{\rm st}(k)$ 
of the PAPD model for $\eta_c(m) < \eta < 1$,
we differentiate the generating function $G(u)$, given by Eq. (\ref{eq:GuHyper}),
according to

\begin{equation}
P_{\rm st}(k) = \frac{1}{k!} \frac{ d^k G(u) }{ d u^k }  \bigg\vert_{u=0}.
\label{eq:PtkD}
\end{equation}

\noindent
Using identity 15.5.6 from Ref. 
\cite{Olver2010},
we obtain the stationary degree distribution

\begin{equation}
P_{\rm st}(k) = \frac{\langle K\rangle}{m\alpha_-(u_+-u_-)}
\left(1-\frac{u_-}{u_+}\right)^{\alpha_+} \frac{(\alpha_++\alpha_-)_k}{(1+\alpha_-)_k}u_+^{-k}
\, _2F_1 \left[ \left.
\begin{array}{c}
1-\alpha_+, \alpha_- \\
\alpha_-+1 + k
\end{array}
\right|    \frac{-u_-}{u_+-u_-}
\right].
\label{eq:Pstk}
\end{equation}

\noindent
Eq. (\ref{eq:Pstk})
is the central analytical result of the paper.
It provides a closed-form expression for the
stationary degree distribution,
in terms of a single parameter,
namely the mean degree $\langle K \rangle$,
which can be obtained from a numerical solution of Eq. (\ref{eq:K_SCE}).
The parameters $u_{\pm}$ and $\alpha_{\pm}$
also depend on $\langle K \rangle$.
Note that Eq. (\ref{eq:Pstk}) is valid only in the regime of $\eta > \eta_c$,
where $\langle K \rangle > 0$.
Moreover, in the limit of
$\eta \rightarrow \eta_c^{+}$ 
(where $\langle K \rangle, \alpha_{\pm} \rightarrow 0$),
at long times $P_t(k) \rightarrow \delta_{k,0}$.
This limit has to be taken carefully because both $\langle K \rangle$
and $\alpha_{-}$ vanish, however the ratio between them remains finite
[see Eq. (\ref{eq:Alphapm})].
Another key observation is that the Pochhammer symbol 
on the right hand side of Eq. (\ref{eq:Pstk}) satisfies 
$(0)_k = \delta_{k,0}$.

In the large $k$ limit, where
$ k \gg \max \left\{ \alpha_-, | \alpha_+ | \right \} $, 
Eq. (\ref{eq:Pstk}) can be approximated by

\begin{equation}
P_{\rm st}(k) \simeq 
\frac{ \langle K \rangle }{ m \alpha_{-} ( u_{+} - u_{-} ) }
\left( 1 - \frac{u_{-} }{u_{+}} \right)^{\alpha_{+}} 
\frac{ \Gamma(1 + \alpha_{-}) }{ \Gamma(\alpha_{+} + \alpha_{-}) }
k^{-(1-\alpha_+)} e^{-\ln (u_+)k},
\label{eq:PstkAsy}
\end{equation}

\noindent 
where $\Gamma(x)$ is the Euler Gamma function \cite{Olver2010}.
The tail of the stationary degree distribution thus takes the form

\begin{equation}
P_{\rm st}(k) \sim k^{-\gamma} e^{-k/k_{\rm cut}},
\label{eq:Pst_kpe}
\end{equation}

\noindent
which is a Gamma distribution
\cite{Johnson1994},
where 
the cutoff scale $k_{\rm cut}$
is given by 

\begin{equation}
k_{\rm cut} = \frac{1}{\ln (u_{+})},
\label{eq:k_cut}
\end{equation}

\noindent
and the exponent $\gamma$ is given by

\begin{equation}
\gamma = 1 - \alpha_{+}.
\label{eq:gamma_1ma}
\end{equation}

\begin{figure}
\begin{center}
\includegraphics[width=5cm]{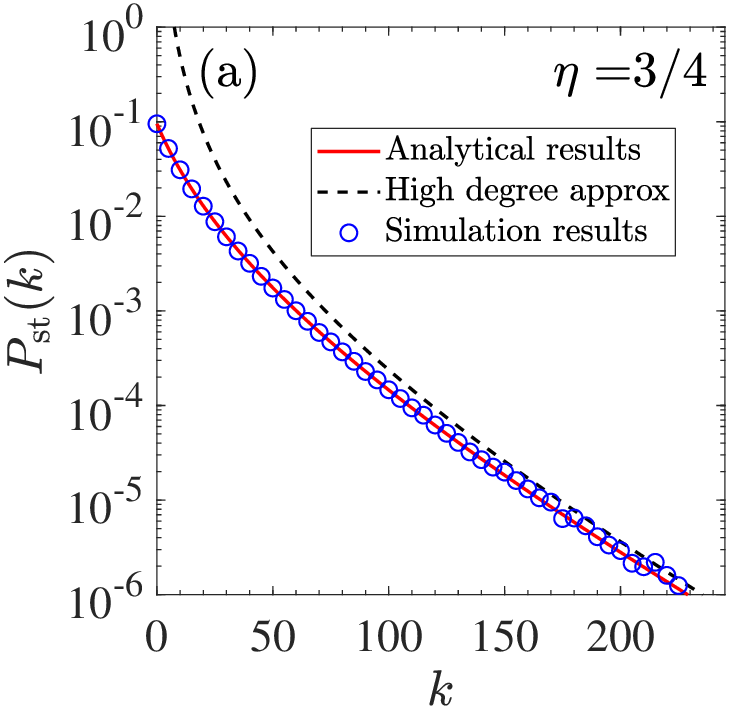} 
\includegraphics[width=5cm]{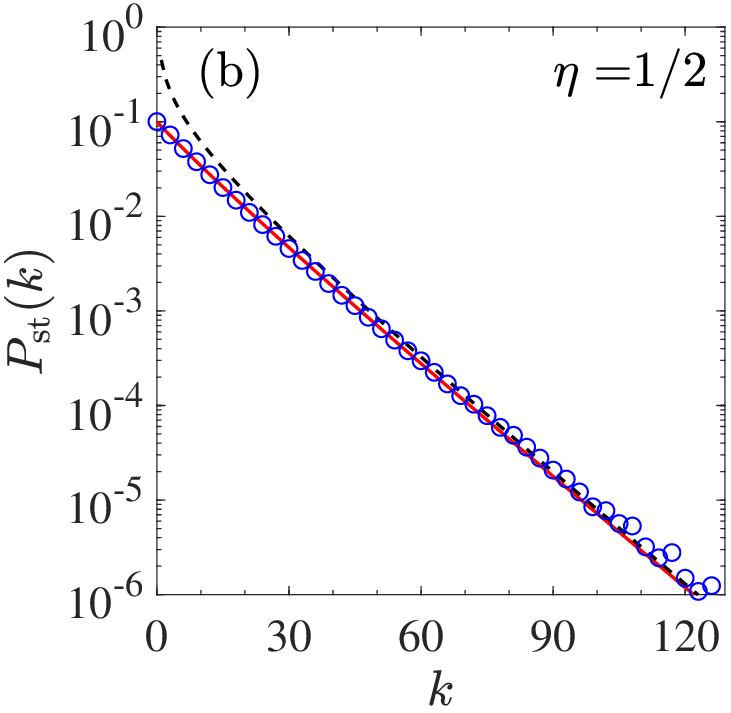}
\includegraphics[width=5cm]{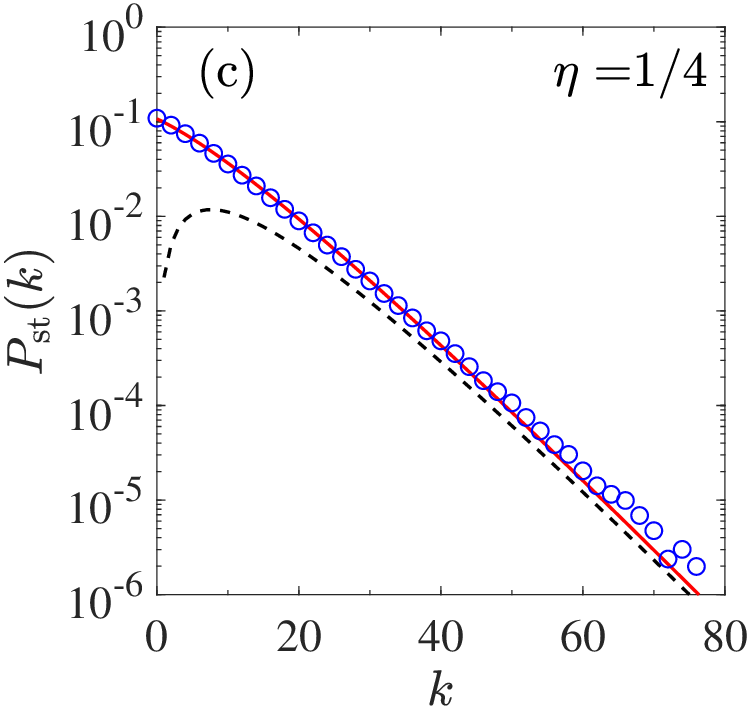}
\end{center}
\caption{
(Color online)  
Analytical results (solid lines)  
for the stationary degree distribution $P_{\rm st}(k)$
of the PAPD model with $m=10$, 
in the regime of overall network growth 
($0 < \eta < 1$),
obtained from Eq. (\ref{eq:Pstk}).
The results are presented for 
(a) $\eta=3/4$,
(b) $\eta=1/2$ 
and
(c) $\eta=1/4$. 
The analytical results are in good agreement with the results
obtained from computer simulations (circles), apart from some
deviations in the tail, which become more pronounced as $\eta$ is decreased.
In the simulations, the initial network  
is an ER network of size $N_0=200$ 
with mean degree $\langle K\rangle_0=3$, 
and the results are shown for networks of size $N_t=20,000$.
We also present the results obtained from the asymptotic expression
(dashed lines)
for the tail of $P_{\rm st}(k)$, 
given by Eq. (\ref{eq:PstkAsy}). 
}
\label{fig:3}
\end{figure}

In Fig.  \ref{fig:3}  we present analytical results (solid lines)
for the stationary
degree distribution $P_{\rm st}(k)$ in the regime of 
overall network growth 
($0 < \eta < 1$) 
with $m=10$,  
obtained from Eq. (\ref{eq:Pstk}). 
Results are presented 
for 
(a) $\eta=3/4$, 
(b) $\eta=1/2$ 
and
(c) $\eta=1/4$. The dashed lines show the asymptotic tails, 
given by Eq. (\ref{eq:PstkAsy}). To validate our calculations, 
we compare them to simulation results (circles), 
and good agreement is found.
Nevertheless, we do observe growing 
discrepancies (though still minor) between them as $\eta$ 
decreases, a matter that will be further addressed later on.

\begin{figure}
\begin{center}
\includegraphics[width=5cm]{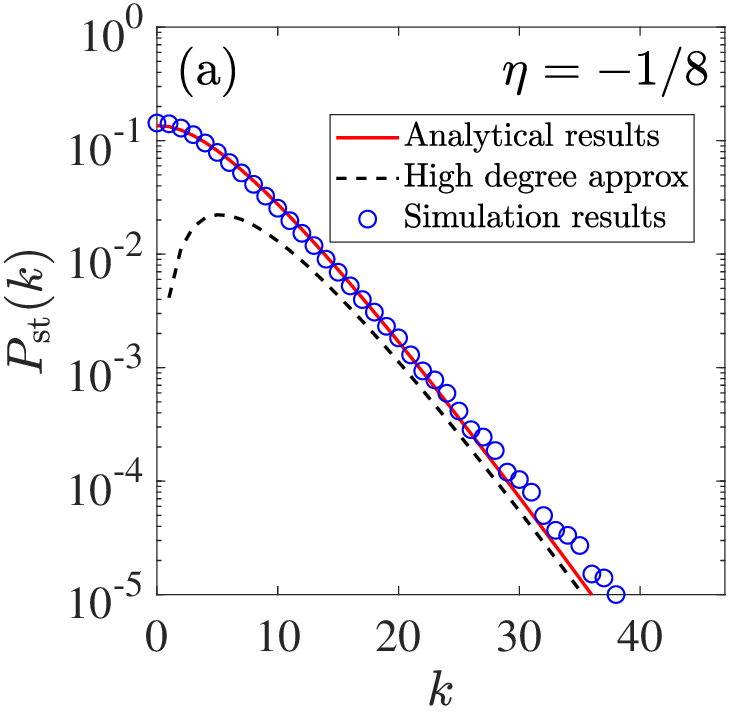} 
\includegraphics[width=5cm]{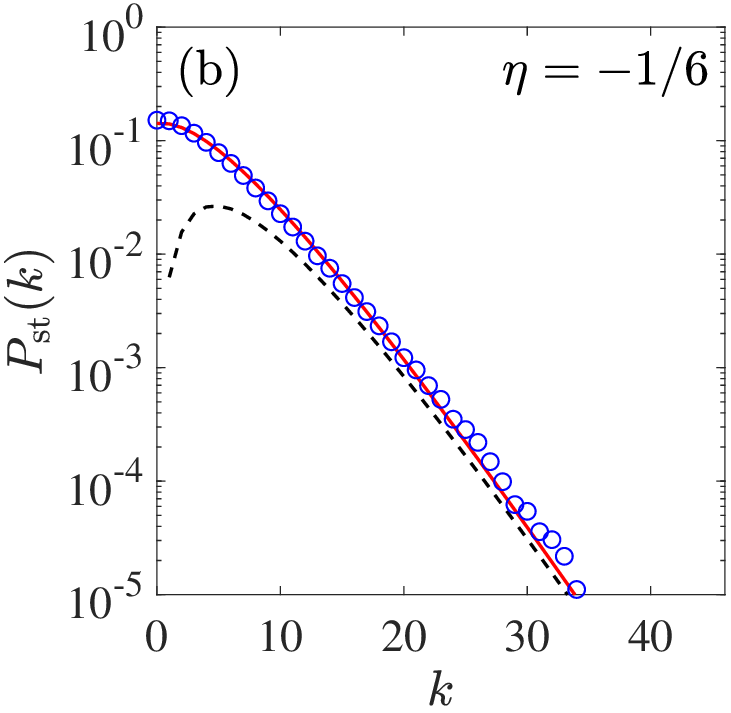}
\includegraphics[width=5cm]{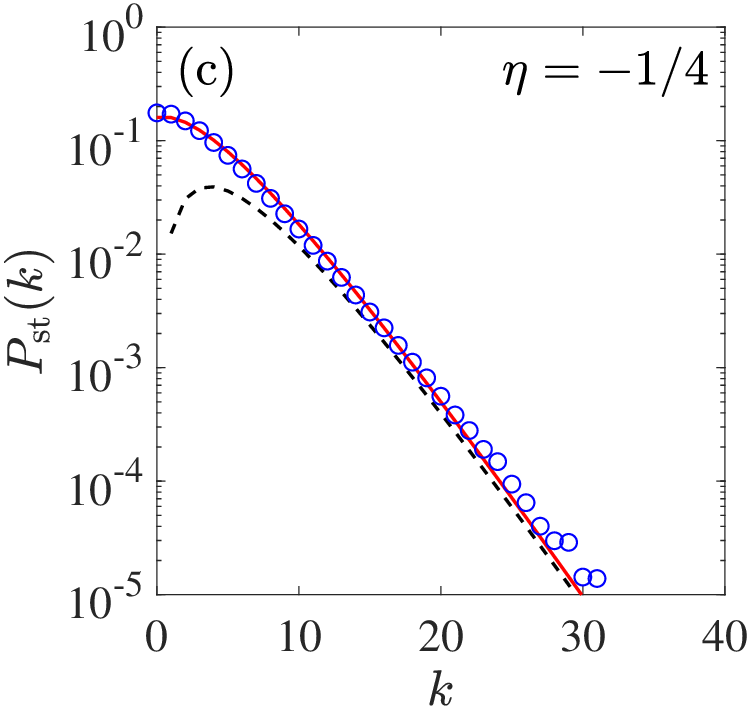}
\end{center}
\caption{
(Color online) 
Analytical results (solid lines)  
for the stationary degree distributions $P_{\rm st}(k)$ 
of the PAPD model with $m=10$,
in the regime of overall network contraction 
($\eta_c < \eta < 0$),
obtained from Eq. (\ref{eq:Pstk}).
The results are presented  
for 
(a) $\eta=-1/8$, 
(b) $\eta=-1/6$ 
and
(c) $\eta=-1/4$. 
The analytical results are in good agreement with the
results obtained from computer simulations (circles), 
apart from some deviations in the tail.
In the simulations, the initial network is an ER
network of size $N_0=20,000$ with mean degree 
$\langle K \rangle_0 = 8$, 
and the results are shown for networks of size $N_t=4,000$.
We also present the results obtained from the asymptotic expression
(dashed lines)
for the tail of $P_{\rm st}(k)$, 
given by Eq. (\ref{eq:PstkAsy}). 
}
\label{fig:4}
\end{figure}

In Fig.  \ref{fig:4}  we present analytical results (solid lines)
for the stationary
degree distribution $P_{\rm st}(k)$ in the regime of overall 
network contraction where 
$\eta_c < \eta <0$,
obtained from Eq. (\ref{eq:Pstk}).
The results are shown for $m=10$, where $\eta_c = -2/3$,  
and for 
(a) $\eta=-1/8$;
(b) $\eta=-1/6$;
and
(c) $\eta=-1/4$. 
The dashed lines show the asymptotic tails, 
given by Eq. (\ref{eq:PstkAsy}).
To validate our calculations, we compare them to simulation results (circles).
As in Fig.  \ref{fig:3}, we find good agreement 
between the analytical results and the corresponding simulation 
results, but with growing discrepancies (though still minor) 
between them as $\eta$ decreases.

\begin{figure}
\begin{center}
\includegraphics[width=7cm]{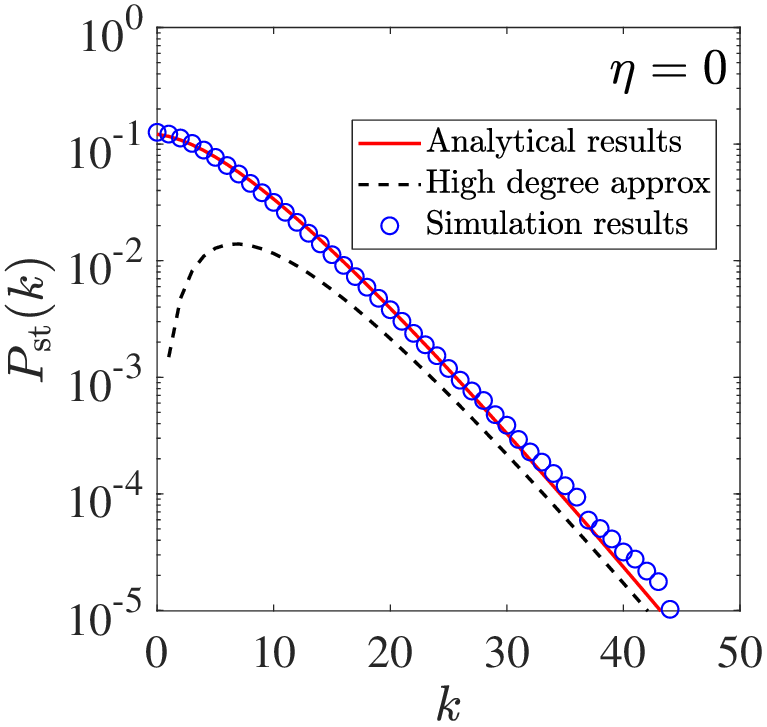}
\end{center}
\caption{
(Color online) 
Analytical results (solid line)
for the stationary
degree distribution $P_{\rm st}(k)$ in the special case of $\eta=0$
with $m=10$,  
obtained from Eq. (\ref{eq:Pstk}). 
In this case, the mean degree $\langle K \rangle$ is evaluated using
the approximated expression of
Eq. (\ref{eq:Kmean_for_eta0}).
In the simulations, the initial network is an ER
network of size $N_0=20,000$ with mean degree 
$\langle K \rangle_0 = 8$, 
and the results are shown at time $t=120,000$.
The dashed line shows the asymptotic tail 
given by Eq. (\ref{eq:PstkAsy}).
}
\label{fig:5}
\end{figure}

In Fig.  \ref{fig:5}  we present analytical results (solid lines)
for the stationary
degree distribution $P_{\rm st}(k)$ in the special case of $\eta=0$
with $m=10$, 
obtained from Eq. (\ref{eq:Pstk}). 
In this case, the value of $\langle K \rangle$ is taken from 
the closed-form approximation given by
Eq. (\ref{eq:Kmean_for_eta0}).
The dashed line shows the asymptotic tail, 
given by Eq. (\ref{eq:PstkAsy}).

In the case of pure growth ($\eta=1$) the stationary degree distribution
is given by
\cite{Budnick2025}

\begin{equation}
P_{\rm st}(k) = 
\frac{4 m (2 m + 1)}{ (k+2m) (k+2m+1) (k+2m+2) }.
\label{eq:Pst_Eta_1}
\end{equation}

\noindent 
In the large $k$ limit, the stationary degree distribution
exhibits a power-law tail of the form 

\begin{equation}
P_{\rm st}(k) \sim k^{-3}.
\label{eq:Pst_km3}
\end{equation}

\noindent
This tail resembles 
the degree distribution of the 
BA model 
\cite{Krapivsky2000,Dorogovtsev2000}.
In this distribution, the mean degree $\langle K \rangle$ is finite,
but the second moment $\langle K^2 \rangle$ diverges logarithmically. 
As shown in Eq. 
(\ref{eq:PstkAsy}), 
even a low rate of 
preferential node deletion breaks down the power-law 
behavior and replaces it by an exponential tail,
characterized by a well-defined scale.
This implies that there is a structural phase transition at $\eta=1$.

At $\eta = 1$, the tail of $P_{\rm st}(k)$ changes smoothly
from a power-law form [Eq. (\ref{eq:Pst_km3})]
into a Gamma distribution
[Eq. (\ref{eq:Pst_kpe})].
Plugging the leading 
contribution 
from Eq. (\ref{eq:K_1}),
in the limit of 
$\eta \rightarrow 1^-$, 
into Eqs. (\ref{eq:upum}) 
and (\ref{eq:Alphapm}), 
namely replacing
$\langle K \rangle$  by $2m$,
we find that 
$u_+ (\eta \rightarrow 1^{-}) \rightarrow 1$ 
and 
$\alpha_+ (\eta \rightarrow 1^{-})  \rightarrow -2$. 
This implies that for 
$\eta \rightarrow 1^-$, 
the exponent obeys
$\gamma \rightarrow 3$ 
and the cutoff scale diverges as
$k_{\rm cut} \rightarrow \infty$, 
recovering the tail of 
$P_{\rm st}(k)$ 
for $\eta=1$,
given by 
Eq. (\ref{eq:Pst_km3}). 
Taking into account sub-leading terms in 
Eq. (\ref{eq:upum}) as $\eta\rightarrow 1^-$,
it is found that 
$k_{\rm cut} = 2m/(1-\eta)+\mathcal{O}(1)$ 
and 
$\gamma=3 - 2 (m+3) ( 1 - \eta ) / m + \mathcal{O} ( 1 - \eta )^2$. 
Note that the divergence of the cutoff scale  
$k_{\rm cut} \sim (1-\eta)^{-1}$ 
as $\eta \rightarrow 1^{-}$ 
resembles the diverging correlation length in
ferromagnetic systems near criticality  
\cite{Pathria2021}.

\begin{figure}
\begin{center}
\includegraphics[width=7cm]{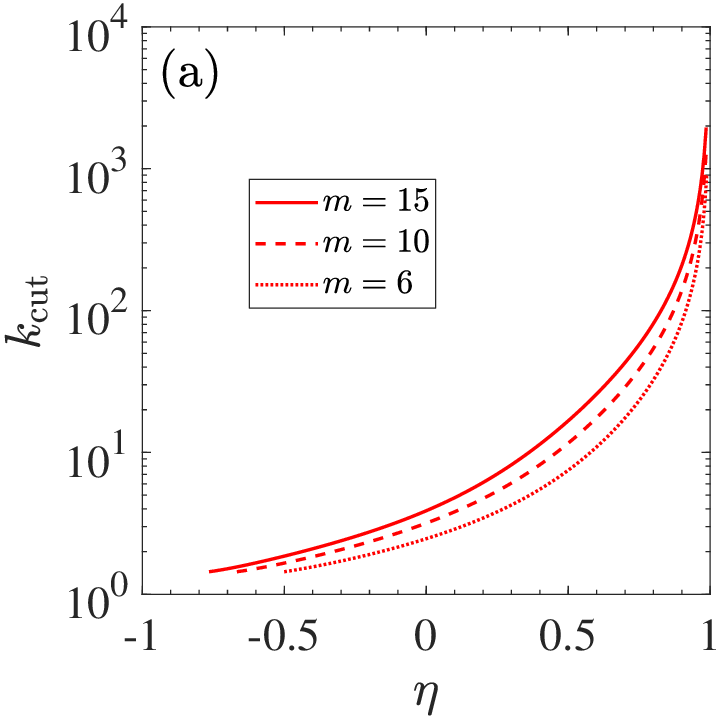} \hspace{0.4 cm}
\includegraphics[width=7cm]{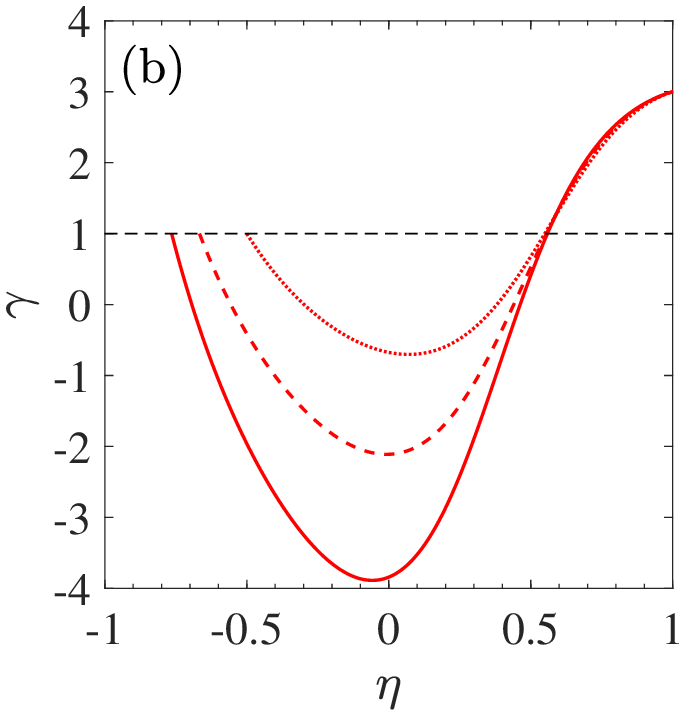}
\end{center}
\caption{
(Color online) 
Analytical results for (a) the cutoff scale $k_{\rm cut}$,
obtained from Eq. (\ref{eq:k_cut}), and 
(b) the exponent $\gamma$, 
obtained from Eq. (\ref{eq:gamma_1ma}), 
as a function of $\eta$, for $m=6$ (dotted lines), $10$ (dashed lines) and $15$ (solid lines).
The cutoff scale $k_{\rm cut}$ and the exponent $\gamma$ characterize
the tail of the stationary degree distribution, 
$P_{\rm st}(k)\sim k^{-\gamma}e^{-k/k_{\rm cut}}$,
given by Eq. (\ref{eq:Pst_kpe}).
}
\label{fig:6}
\end{figure}

In Fig. \ref{fig:6} we present analytical results for 
(a) the cutoff scale $k_{\rm cut}$, obtained from Eq. (\ref{eq:k_cut}),
and (b) the exponent $\gamma$, obtained from Eq. (\ref{eq:gamma_1ma}),
as a function of $\eta$, 
for $m=6$ (dotted lines), $10$ (dashed lines) and $15$ (solid lines).
The analytical results are evaluated from a numerical solution of 
Eq. (\ref{eq:K_SCE}) for the mean degree $\langle K\rangle$, 
which is inserted into 
Eqs. (\ref{eq:Alphapm}) and (\ref{eq:upum}).

\begin{figure}
\begin{center}
\includegraphics[width=7cm]{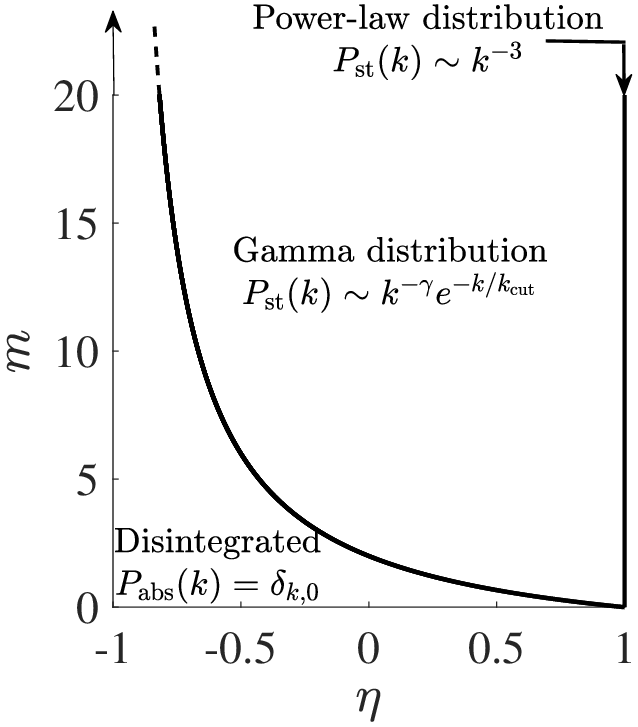} \hspace{0.4 cm}
\end{center}
\caption{
The phase diagram  
of the PAPD model in terms of the growth/contraction rate $\eta$ and 
the parameter $m$.
The boundary line $m_c(\eta) = 2 (1-\eta)/(1+\eta)$
[Eq. (\ref{eq:m_c_eta})]
separates 
between a phase that exhibits a stationary degree distribution
$P_{\rm st}(k)$ that follows the Gamma distribution
and a phase in which the network quickly disintegrates
into an
absorbing state, 
characterized by the null degree distribution $P_{\rm abs}(k)=\delta_{k,0}$. 
The boundary line at $\eta=1$ corresponds to the limit of pure network growth, 
which is
characterized by a degree distribution that exhibits a power-law tail of the form 
$P_{\rm st}(k) \sim k^{-3}$,
implying a structural phase transition at $\eta=1$.
}
\label{fig:7}
\end{figure}

In Fig. \ref{fig:7}  we present the 
phase diagram of networks that evolve according to the PAPD model. 
For each value of $m$ there is a critical value 
$\eta_c(m)$ of the growth/contraction rate, given by Eq. (\ref{eq:eta_c}).
For $\eta > \eta_c(m)$ there is a stationary phase with
degree distribution which exhibits a tail of the form
$P_{\rm st}(k) \sim k^{-\gamma} e^{-k/k_{\rm cut}}$.
In contrast, for $\eta < \eta_c(m)$ there is no stationary 
state. In this regime, the network disintegrates
and converges into an absorbing state that consists of a set
of isolated nodes, namely
$P_{\rm abs}(k)=\delta_{k,0}$,
where $\delta_{i,j}$ is the Kronecker delta.
In the case of pure network growth ($\eta=1$), the stationary degree
distribution exhibits a power-law tail of the form
$P_{\rm st}(k) \sim k^{-3}$,
implying a structural phyase transition at $\eta=1$.

\section{Discussion}

Unlike the common approach of solving the master equation using direct numerical integration,
in this paper we present an analytical solution of the master equation for the stationary state
of the degree distribution $P_{\rm st}(k)$.
Analytical solutions of such master equations were obtained 
for other models of network evolution
such as RARD model 
\cite{Budnick2022},
the PARD model
\cite{Budnick2025}
and node-duplication models
\cite{Lambiotte2016,Bhat2016,Steinbock2019}.
These equations account for the distribution $P_t(k)$ but do not account for
the correlations between the degrees of adjacent nodes.
In order to account for degree-degree correlations, one needs to consider
a higher-dimensional master equation for the joint degree distribution
$P_t(k,k')$ of the degrees of pairs of adjacent nodes.
In this case, the number of coupled equations is much larger and
are more difficult to solve analytically.

Since the master equation 
(\ref{eq:dP(t)/dtRC0}) 
ignores degree-degree correlations
[as can be seen in Eq. (\ref{eq:Pkk'Approx})]
we analyze these correlations
using computer simulations.
To this end, we calculate
the correlation coefficient
\cite{Newman2002,Newman2003}

\begin{equation}
\rho = \frac{ \langle K K'\rangle -
\langle  \widetilde K  \rangle^2 }
{\langle {\widetilde K}^2 \rangle - 
\langle {\widetilde K} \rangle^2},
\label{eq:CorrCoeff}
\end{equation}

\noindent 
where

\begin{equation}
\langle K K' \rangle = \sum_{k=1}^{\infty} \sum_{k'=1}^{\infty} k k' P_{\rm st}(k,k')
\label{eq:msm}
\end{equation}

\noindent
is the mixed second moment of the joint stationary degree 
distribution $P_{\rm st}(k,k')$ of pairs of adjacent nodes,
while 
$\langle \widetilde K \rangle$ 
and 
$\langle \widetilde K^2 \rangle$
are the first and second moments of the
degree distribution

\begin{equation}
\widetilde P_{\rm st}(k) = \frac{k}{\langle K \rangle} P_{\rm st}(k),
\end{equation}

\noindent
of nodes selected via a random edge from the stationary degree distribution
$P_{\rm st}(k)$.
Note that within the framework of the master equation
(\ref{eq:dP(t)/dtRC0}) the degree-degree correlations are neglected, such that
the joint degree distribution is approximated by
$P_{\rm st}(k,k') = \widetilde P(k) \widetilde P(k')$ and
the mixed second moment is approximated by
$\langle K K' \rangle = \langle \widetilde K \rangle^2$.
As a result, within this framework the correlation coefficient is
$\rho=0$.

\begin{figure}
\begin{center}
\includegraphics[width=7cm]{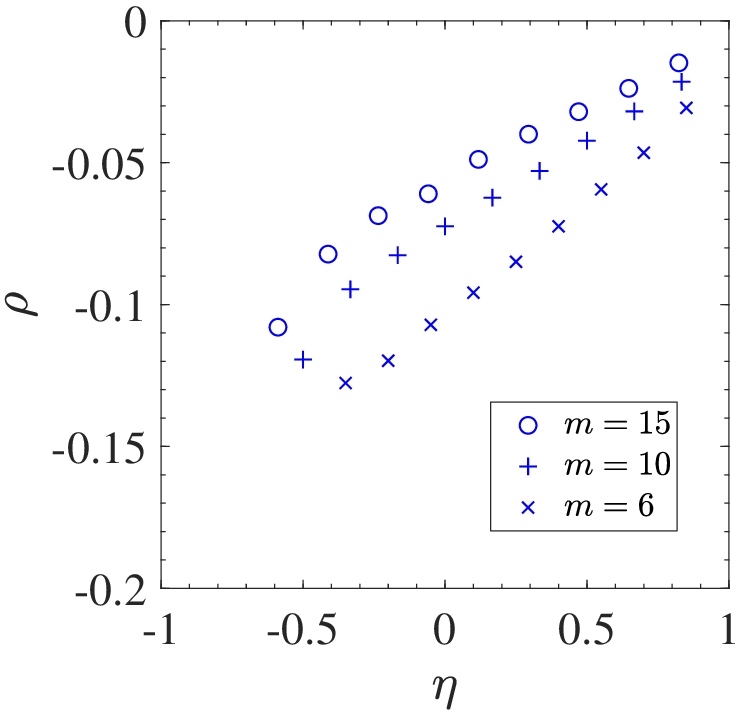}
\end{center}
\caption{
(Color online) 
Simulation results
for correlation coefficient $\rho$  
[Eq. (\ref{eq:CorrCoeff})]
between the degrees of pairs of adjacent nodes
in the PAPD model 
as a function of the growth/contraction rate $\eta$.
The results are shown for $m=6$, $10$ and $15$
(bottom to top).
It is found that $\rho$ is negative, which implies that
the network is disassortative.
Moreover, the correlations become stronger as $\eta$ is decreased.
For $0 \le \eta < 1$ the parameters of the simulations are the same as
in Fig. \ref{fig:3}, while for $-1 < \eta < 0$ the parameters are the same
as in Fig. \ref{fig:4}.
}
\label{fig:8}
\end{figure}

In Fig. \ref{fig:8} we present simulation results for the 
correlation coefficient $\rho$,
as a function of the growth/contraction rate $\eta$,
for $m=6$, $10$ and $15$ (bottom to top). 
It is found that the correlation coefficient is negative
for all three values of $m$. 
This implies that the PAPD network is disassortative,
namely high-degree nodes tend to be connected to low-degree
nodes and vice versa
\cite{Newman2002,Newman2003}.
The disassortativity may be attributed to the asymmetry of 
the edge-addition step, in which a node selected uniformly at
random is connected to a node selected preferentially,
namely with probability proportional to its degree.
As a result, in each edge the preferentially selected end-node is
expected to be of a higher degree than the node selected uniformly at random.
In particular, this edge-addition cannot generate dimers 
(where both end-nodes are of degree $k=1$).
As a result, isolated nodes (of degree $k=0$) must connect to nodes
of degree $k \ge 1$, resulting in a bond between a node of degree
$k=1$ and a node of degree $k \ge 2$.
Fig. \ref{fig:8} clearly shows that the absolute value of the
correlation coefficient $\rho$ increases as $\eta$ is decreased,
namely the correlations are weak in the limit
of pure growth ($\eta=1$) and become stronger as $\eta$
is decreased towards $\eta_c(m)$.
This is consistent with the growing discrepancy in the tail of 
$P_{\rm st}(k)$ when $\eta$ is decreased,
as can be seen in Figs. \ref{fig:3} and \ref{fig:4}.

Beyond the degree distribution, it is interesting to explore the large scale structure
of the network in the stationary state.
Percolation theory provides a useful framework for such analysis
\cite{Havlin2010,Newman2010}.
In particular, it provides tools to distinguish between sub-critical networks,
which consist of finite tree components, and super-critical networks, which
exhibit a coexistence between a giant component and finite tree components.
The Molloy-Reed criterion states that  
for a class of random networks referred to as configuration model networks,
in the super-critical regime,
the condition 
\cite{Molloy1995,Molloy1998}

\begin{equation}
\frac{ \langle K^2 \rangle}{ \langle K \rangle } > 2 
\end{equation}

\noindent
is satisfied.
Configuration model networks are random networks in which
the degree sequence is drawn from
a given degree distribution $P(k)$, while the connectivity
is random 
\cite{Bollobas1984,Newman2001}.
These networks form
a maximum entropy ensemble, under the 
constraint imposed by the degree distribution.
As a result, in configuration model networks the
degrees of adjacent nodes are uncorrelated.

Since the PAPD network exhibits degree-degree correlations,
it clearly does not belong to the class of configuration model networks.
In spite of this observation, it would be interesting to see what we can 
learn from applying the Molloy-Reed condition to these networks.
Using Eq. (\ref{eq:KK^2}), it is found that the PAPD network satisfies

\begin{equation}
\frac{ \langle K^2 \rangle}{ \langle K \rangle } = r m - \frac{r - 1}{2} \langle K \rangle.
\end{equation}

\noindent
Therefore, the Molloy-Reed criterion takes the form

\begin{equation}
r m - \frac{r - 1}{2} \langle K \rangle > 2.
\end{equation}

\noindent
This inequality should be solved separately for $\eta > 0$ and for $\eta < 0$.
For $\eta<0$, we obtain the condition

\begin{equation}
\langle K \rangle > - 2 m \frac{ r - \frac{2}{m} }{ | 1 - r | },
\label{eq:Km_etaneg}
\end{equation}

\noindent
where $|x|$ is the absolute value of $x$.
This implies that for $r > r_c(m) = 2/m$,
the right hand side of Eq. (\ref{eq:Km_etaneg}) 
is negative and the inequality must be satisfied.
We thus conclude that
the stationary state solution $P_{\rm st}(k)$
corresponds to a supercritical network, which exhibits a giant component.
In case that $\eta > 0$ ($r > 1$) we obtain the condition

\begin{equation}
\langle K \rangle < 2 m \frac{ r - \frac{2}{m} }{  r - 1  }.
\label{eq:Km1}
\end{equation}

\noindent
In case that $m \ge 2$, the ratio on the right hand side of Eq. (\ref{eq:Km1})
satisfies

\begin{equation}
\frac{r - \frac{2}{m}}{r - 1} > 1.
\end{equation}

\noindent
The mean degree satisfies 
$\langle K \rangle = 2 m$
for $\eta = 1$
and 
$\langle K \rangle < 2 m$
for $\eta < 1$.
As a result, Eq. (\ref{eq:Km1}) must be satisfied in case of $m \ge 2$.
For the case of $m < 2$ we tested Eq. (\ref{eq:Km1}) numerically
and confirmed that it is indeed valid for 
$r > r_c(m)$.

The results presented above indicate that at the level of the 
(correlation-ignoring) master equation,
the boundary between the stationary phase and the disintegrated phase
coincides with the percolation transition.
This implies that the stationary phase consists of super-critical networks
in which there is a giant component that coexists with finite tree components.
Using computer simulations we confirmed that in the stationary phase the
networks exhibit a giant component and are thus super-critical.
However, in the simulation results it is difficult to find the precise
location of the percolation threshold.
This may be due to the degree-degree correlations, which attain their
strongest values just above $\eta = \eta_c(m)$.
As a result, the correlations may affect the behavior near the percolation threshold.

It would be interesting to explore the percolation properties of the
PAPD networks in their stationary state, calculating the 
fraction of nodes that belong to the giant component and
the distribution of the sizes of the finite components
\cite{Newman2007}.
As a starting point, one can use methodologies developed
for configuration model networks
\cite{Tishby2018}.
It would also be interesting to explore the large scale structure
of these networks, characterized by the distribution of shortest path lengths (DSPL)
between pairs of random nodes
\cite{Dorogovtsev2003,Katzav2015,Nitzan2016,Steinbock2017}.

\section{Summary}

In this paper we studied the effect of preferential node deletion on
the structure of networks that evolve via node addition and preferential attachment.
To this end, we introduced the PAPD model,
in which at each time step, with probability $P_{\rm add}=(1+\eta)/2$ there is a growth step where
an isolated node is added to the network, followed by the addition of $m$ 
edges, where each edge connects a node selected uniformly at random to a node
selected preferentially in proportion to its degree.
Alternatively, with probability $P_{\rm del}=(1-\eta)/2$ there is a contraction step,
in which a preferentially selected node is deleted and its links are erased. 
For $0 < \eta \le 1$ the overall process is of network growth,
while for $-1 \le \eta < 0$ the overall process is of network contraction.

Using the master equation and computer simulations,
we explored the time-dependent degree distribution $P_t(k)$ 
of the PAPD network.
It was found that for 
each value of $m>0$ there is a critical value 
$\eta_c(m) = - (m-2)/(m+2)$
such that for
$\eta_c(m) < \eta \le 1$ 
the degree distribution $P_t(k)$ converges towards a stationary distribution
$P_{\rm st}(k)$.
In the special case of pure growth, where $\eta=1$, 
the stationary degree distribution  $P_{\rm st}(k)$
exhibits a power-law tail, which is a characteristic of scale-free networks.
In contrast, for $\eta_c(m) < \eta < 1$ the distribution $P_{\rm st}(k)$
exhibits an exponential tail, which has a well-defined scale.
This implies that there is a phase transition at $\eta = 1$.
These results illustrate the enhanced sensitivity of evolving networks  to
preferential node deletion, in comparison to random node deletion
\cite{Budnick2022,Budnick2025}.

While for $\eta \ge \max \left\{ \eta_c(m),0 \right \}$ 
the stationary degree distribution $P_{\rm st}(k)$ lasts indefinitely, 
for $\eta_c(m) < \eta < 0$ (and $m > 2$) 
it persists for a finite lifetime, until the network vanishes.
It was also found that in the regime of 
$-1 \le \eta \le  \eta_c(m)$ 
the time-dependent degree distribution
$P_t(k)$ does not converge towards a stationary form, but continues to
evolve until the network is reduced to a set of isolated nodes.
The behavior in this regime resembles the case of pure preferential
deletion, considered in Refs.
\cite{Tishby2019,Tishby2020}.

The results presented above provide insight on the structure of transient social networks,
such as dating networks and job-seeking platforms,
in which the turnover is intrinsically high. 
Unlike permanent social networks, which tend to exhibit a scale-free structure,
which is characterized by power-law degree distribution,
the degree distributions of transient networks are expected to 
follow a Gamma distribution, which exhibits an exponential tail
with a well defined scale.

\appendix

\section{Proof that $\alpha_->0$}

Consider the generating function $G(u)$ of the stationary degree distribution $P_{\rm st}(k)$.
It satisfies Eq. (\ref{eq:diffGu}), which can be expressed in the form

\begin{equation}
(u-u_-)(u_+-u) \frac{ d G(u) }{d u} = 
\langle K\rangle
\left[ u-\left(1+\frac{r-1}{rm}\right)\right]G(u)+\frac{\langle K\rangle}{m}.
\label{eq:diffGuRegular}
\end{equation}

\noindent
Substituting $u=u_-$ into Eq. (\ref{eq:diffGuRegular}) and using the 
definition of $\alpha_-$ given in Eq. (\ref{eq:Alphapm}) we get

\begin{equation}
G(u_-) = \frac{\langle K\rangle}{m\alpha_-(u_+-u_-)}.
\label{eq:Gum}
\end{equation}

\noindent
Since $0<u_-<1<u_+$  
and from the definition of $G(u)$ [Eq. (\ref{eq:GuDef})], it is clear that 
one must also have $0 \le G(u_-) \le 1$. 
This rules out the possibility $\alpha_-\leq 0$. 
Therefore, $\alpha_{-}$ must be positive.

\begin{figure}
\begin{center}
\includegraphics[width=7cm]{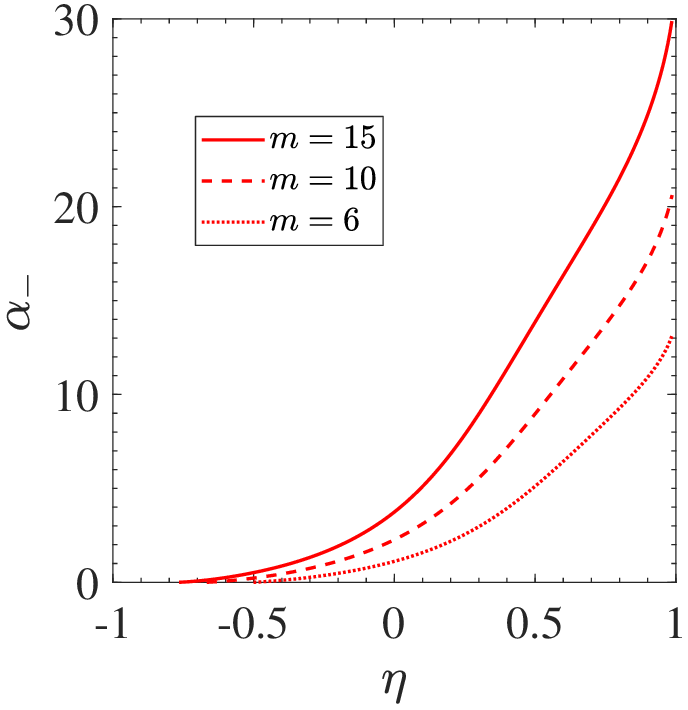} \hspace{0.4 cm}
\end{center}
\caption{
(Color online)
The parameter $\alpha_{-}$ as a function of the growth contraction 
rate $\eta$ for $m=6$ (dotted line), $m=10$ (dashed line) and $m=15$ (solid line),
obtained from Eq. (\ref{eq:Alphapm}).
For each value of $m$, the parameter $\alpha_{-}=0$ at $\eta=\eta_c(m)$,
and increases monotonically as a function of $\eta$ for $\eta_c(m) < \eta < 1$,
approaching $\alpha_{-} = 2 m + 2$ as $\eta \rightarrow 1$.
}
\label{fig:9}
\end{figure}

In Fig. \ref{fig:9} we present the parameter $\alpha_{-}$ as a function of $\eta$,
obtained from Eq. (\ref{eq:Alphapm}),
for $m=6$, $10$ and $15$ (bottom to top).
For each value of $m$, the parameter $\alpha_{-}=0$ at $\eta=\eta_c(m)$,
and increases monotonically as a function of $\eta$ for $\eta_c(m) < \eta < 1$.
In the limit of $\eta \rightarrow 1$ ($r \rightarrow \infty$),
the mean degree $\langle K \rangle \rightarrow 2m$
and $\alpha_{-} \rightarrow 2 m + 2$.

\section{Approximations for the mean degree $\langle K\rangle$ near criticality}

In this Appendix we derive approximate formulae for the mean degree $\langle K \rangle$
of the stationary degree distribution $P_{\rm st}(k)$ in the limits of 
$\eta \rightarrow \eta_c^{+}$
and
$\eta \rightarrow 1^{-}$.

\subsection{The mean degree $\langle K \rangle$ in the limit of $\eta \rightarrow \eta_c^{+}(m)$}

To approximate $\langle K\rangle$ as $\eta\rightarrow \eta_c^{+}(m)$ we expand 
Eq. (\ref{eq:K_SCE}) around $(\langle K\rangle,r,m)=(0,2/m,m)$. 
The only non-trivial part of the analysis is the expansion of the hypergeometric function. 
In order to carry out this expansion, we first expand
$u_\pm$ and $\alpha_\pm$ up to 
$\mathcal{O} [ \eta - \eta_c(m) ]$ and
$\mathcal{O} (\langle K \rangle)$.
Using  
Eq. (\ref{eq:2F1}) 
we obtain

\begin{eqnarray}
\, _2F_1 \left[ \left.
\begin{array}{c}
1-\alpha_+, \alpha_- \\
\alpha_-+1
\end{array}
\right|    \frac{1-u_-}{u_+-u_-}
\right] 
\simeq  
\ \ \ \ \ \ \ \ \ \ \ \ \ \ \ \ \ \ \ \ \ \ \ \ \ \ \ \ \ \ \ \ \ \ \ \ \ \ \ \ \ \ \ \ \ \    
\nonumber \\
\sum_{n=0}^{\infty}
\frac{\left[ 1 - \left(1 - \frac{2}{3m} \right)
\langle K\rangle \right]_n 
\left(\frac{2\langle K\rangle}{3m}\right)_n}
{\left(\frac{2\langle K\rangle}{3m}+1\right)_n}
\frac{\left[ \frac{1}{3}-\frac{2\left(m-2\right)}{27m}\langle K\rangle
+\frac{m}{27}\left(r-\frac{2}{m}\right)\right]^n}{n!}.
\label{eq:hg1}
\end{eqnarray}

\noindent
Expanding the right hand side of Eq. (\ref{eq:hg1}) up to
$\mathcal{O}(\langle K \rangle,r-2/m)$,
we obtain

\begin{equation}
\, _2F_1 \left[ \left.
\begin{array}{c}
1-\alpha_+, \alpha_- \\
\alpha_-+1
\end{array}
\right|    \frac{1-u_-}{u_+-u_-}
\right] 
\simeq 
1+\frac{2\langle K\rangle}{3m}\sum_{n=1}^\infty \frac{1}{n}
\left(\frac{1}{3}\right)^n.
\label{eq:hg2}
\end{equation}

\noindent
Carrying out the summation on the right hand side of Eq. (\ref{eq:hg2}),
we obtain

\begin{equation}
\, _2F_1 \left[ \left.
\begin{array}{c}
1-\alpha_+, \alpha_- \\
\alpha_-+1
\end{array}
\right|    \frac{1-u_-}{u_+-u_-}
\right] 
\simeq
1+\frac{2}{3m}\ln\left(\frac{3}{2}\right)\langle K\rangle.
\label{eq:hg3}
\end{equation}

\noindent
Expanding the pre-factor of the hypergeometric function in Eq. (\ref{eq:K_SCE}) yields

\begin{eqnarray}
\frac{\langle K\rangle}{m\alpha_-(u_+-u_-)}\left(\frac{u_+-u_-}{u_+-1}\right)^{\alpha_+}
\simeq 
\ \ \ \ \ \ \ \ \ \ \ \ \ \ \ \ \ \ \ \ \ \ \ \ \ \ \ \ \ \ \ \ \ \ 
\nonumber \\
1 
+ 
\frac{m\left[m-2+12\ln\left(\frac{3}{2}\right)\right]
-8\ln\left(\frac{3}{2}\right)}{12m}\langle K\rangle
-
\frac{m^2}{16}\left(r-\frac{2}{m}\right).
\label{eq:Rest_Approx_etac}
\end{eqnarray}

\noindent
Plugging in the right hand sides of 
Eqs. (\ref{eq:hg3}) 
and 
(\ref{eq:Rest_Approx_etac}) 
into Eq. (\ref{eq:K_SCE}) 
and expanding up to order
$\mathcal{O} ( \langle K \rangle,r-2/m )$ 
yields

\begin{equation}
1+\left[ \frac{m-2}{12} +\ln\left(\frac{3}{2}\right) \right]
\langle K\rangle -\frac{m^2}{6}\left(r-\frac{2}{m}\right)
+\mathcal{O} [ \eta - \eta_c(m) ]
=1.
\end{equation}

\noindent 
Solving for $\langle K\rangle$ and using 
Eq. (\ref{eq:r}), it is found that
as 
$\eta \rightarrow \eta_c^{+}(m)$,
the mean degree can be approximated by

\begin{equation}
\langle K\rangle 
= 
\frac{(m+2)^2}{m-2+12\ln\left(3/2\right)} [ \eta-\eta_c(m) ] 
+ \mathcal{O} [ \eta-\eta_c(m) ]^2.
\end{equation}

\subsection{The mean degree $\langle K \rangle$ in the limit of $\eta \rightarrow 1^{-}$}

This case is more involved than the previous one. This is due to the fact that in the limit of 
$r \rightarrow \infty$ the left hand side of Eq. (\ref{eq:K_SCE}) 
contains a product of a vanishing expression and a diverging expression.
The vanishing expression is

\begin{equation}
\lim_{r\rightarrow\infty} \left(\frac{u_+-u_-}{u_+-1}\right)^{\alpha_+} = 0
\end{equation}

\noindent 
and the diverging expression is

\begin{equation}
\lim_{r\rightarrow\infty} \, _2F_1 \left[ \left.
\begin{array}{c}
1-\alpha_+, \alpha_- \\
\alpha_-+1
\end{array}
\right|    \frac{1-u_-}{u_+-u_-}
\right] = \infty.
\end{equation}

\noindent 
One way around this is to combine identities 15.8.4 and 5.5.3 
in Ref. 
\cite{Olver2010} 
and transform 
Eq. (\ref{eq:K_SCE}) 
into

\begin{eqnarray}
\frac{\langle K\rangle}{m\alpha_-(u_+-u_-)}
\frac{\Gamma(\alpha_-+1)\Gamma(\alpha_+)}{\Gamma(\alpha_-+\alpha_+)}
\left(\frac{u_+-u_-}{u_+-1}\right)^{\alpha_+}
\, _2F_1 \left[ \left.
\begin{array}{c}
1-\alpha_+, \alpha_- \\
1-\alpha_+
\end{array}
\right|    \frac{u_+-1}{u_+-u_-}
\right] 
\nonumber \\
-
\frac{\langle K\rangle}{m\alpha_{+} (u_+-u_-)}
\, _2F_1 \left[ \left.
\begin{array}{c}
\alpha_-+\alpha_+, 1 \\
\alpha_++1
\end{array}
\right|    \frac{u_+-1}{u_+-u_-}
\right]
= 1.
\label{eq:K_SC_eta1}
\end{eqnarray}

\noindent
Our next step is to expand 
Eq. (\ref{eq:K_SC_eta1}) 
around 
$(\langle K\rangle,r,m) =(2m,\infty,m)$. 
In contrast to the analysis provided in 
Eq. (\ref{eq:hg1}), 
the series representing the two 
hypergeometric functions in Eq. (\ref{eq:K_SC_eta1}) are arranged in 
powers of $1/r$. This arrangement results in the series 
terminating after a finite number of terms based on the specified order of expansion. 
Accordingly, expanding the left hand side of Eq. (\ref{eq:K_SC_eta1}) up to 
$\mathcal{O}(1/r^2)$ we obtain

\begin{eqnarray}
1+\left(1-\frac{2m}{\langle K\rangle}\right)  
\frac{1}{r} +
\left\{ 1 
- \frac{2m ( 3 \langle K \rangle-2 )   }{\langle K\rangle^2} 
- \frac{ 2m [ \langle K \rangle  ( (5 \langle K \rangle+12 )+8 ] }
{ \langle K \rangle^3 (\langle K \rangle +2) }  
\right.
\nonumber \\
\left.
- 
\frac{ 16 m^2  (\langle K \rangle +1 ) }{\langle K \rangle^3}
\left[ H_{ \langle K \rangle + 2}
+\ln\left( \frac{4m}{\langle K\rangle^2 r} \right)  \right]
\right\} 
\frac{1}{r^2}
+\mathcal{O} \left(  \frac{1}{r^3} \right)=1,
\label{eq:K_SC_expan_eta1}
\end{eqnarray}

\noindent 
where $H_n$ is the $n$th harmonic number  
\cite{Olver2010}.
Hence, the sub-leading term of 
$\langle K\rangle$ as $r\rightarrow\infty$ contains a logarithmic correction. 
To satisfy Eq. (\ref{eq:K_SC_expan_eta1}) up to order
$\mathcal{O}(1/r^2)$ 
we use the ansatz

\begin{equation}
\langle K\rangle = 2m+\frac{c_1+c_2\ln r}{r},
\label{eq:Anzats_eta1}
\end{equation}

\noindent 
where $c_1$ and $c_2$ are parameters.
Inserting this ansatz
into Eq. (\ref{eq:K_SC_expan_eta1}), solving for $c_1$ 
and $c_2$ up to order
$\mathcal{O}(1/r^2)$ 
and using 
Eq. (\ref{eq:r}), 
we obtain

\begin{equation}
\langle K\rangle = 2m - \left\{ 2(2m+1)
\left[\ln\left( \frac{2m}{1-\eta} \right) - H_{2m+2} \right]
-7m-\frac{1}{m+1} \right\}\left(1-\eta\right) 
+ \mathcal{O}(1-\eta)^2.
\end{equation}

\section{Approximation for the mean degree $\langle K \rangle$ at $\eta = 0$}

In this Appendix we derive an approximate closed-form expression for the
mean degree $\langle K \rangle$ in the special case of $\eta=0$ ($r=1$).
For $\eta=0$  
the parameters $u_{\pm}$,
given by Eq. (\ref{eq:upum}),
are reduced to  

\begin{equation}
u_{\pm} = \frac{ 1 + 2 m \pm \sqrt{1+4m} }{ 2m }.
\label{eq:upum_eta0}
\end{equation}

\noindent
As a result, the parameters $\alpha_{\pm}$,
given by Eq. (\ref{eq:Alphapm}),
are reduced to

\begin{equation}
\alpha_{\pm} = \frac{ \langle K \rangle }{ 2 }
\left[ 1 \pm \frac{ 1 }{ \sqrt{ 1 + 4 m } } \right].
\label{eq:alphapm_eta0}
\end{equation}

\noindent
These simplified expressions enable us to express the mean degree
$\langle K \rangle$ as a function of $m$.
As a first step, we verify that for $m=1$,
the solution of Eq. (\ref{eq:K_SCE}) yields $\langle K \rangle(m=1,\eta=0) = - 1$.
This solution, in which the mean degree is negative, is clearly out of the physical range.
However, it is useful for the derivation below.
From Appendix B, we know that for $\eta=0$ and $m=2$ the mean degree
is $\langle K \rangle(m=2,\eta=0) = 0$.
Moreover, in the vicinity of $m=2$, one obtains the leading order behavior

\begin{equation}
\langle K \rangle(m,\eta=0) = \frac{1}{ 3 \ln (3/2) } (m-2) + \mathcal{O} \left[ (m-2)^2 \right].
\end{equation}

\noindent
Solving Eq. (\ref{eq:K_SCE}) numerically in the limit of $m \rightarrow \infty$, it is found
that in this limit

\begin{equation}
\langle K \rangle(m \rightarrow \infty,\eta=0) = \frac{2}{\pi} m + \mathcal{O} (1).
\end{equation}

\noindent
Putting together all these constraints, 
it is found that the simplest rational approximation for $\langle K \rangle$ that is
consistent with all of them is given by

\begin{equation}
\langle K \rangle(m,\eta=0) \simeq (m-2) \frac{ a_1 + a_2 (m-2) }{ 1 + a_3 (m-2) },
\label{eq:Kmean_for_eta0}
\end{equation}

\noindent
where

\begin{eqnarray}
a_1 &=& \frac{1}{ 3 \ln (3/2) }
\nonumber \\
a_2 &=& \frac{2}{\pi - 2} (1 - a_1)
\nonumber \\
a_3 &=& \frac{\pi}{\pi - 2} (1 - a_1).
\end{eqnarray}

\noindent
Thus, in the special case of $\eta=0$,
Eq. (\ref{eq:Kmean_for_eta0}) provides a
highly accurate closed-form expression for the stationary
mean degree $\langle K \rangle(m,\eta=0)$.
Inserting $\langle K \rangle$ from Eq. (\ref{eq:Kmean_for_eta0})
into the right hand side of Eq. (\ref{eq:VarK}),
one can also obtain a closed-form approximation for the variance ${\rm Var}(K)$.

\begin{figure}
\begin{center}
\includegraphics[width=7cm]{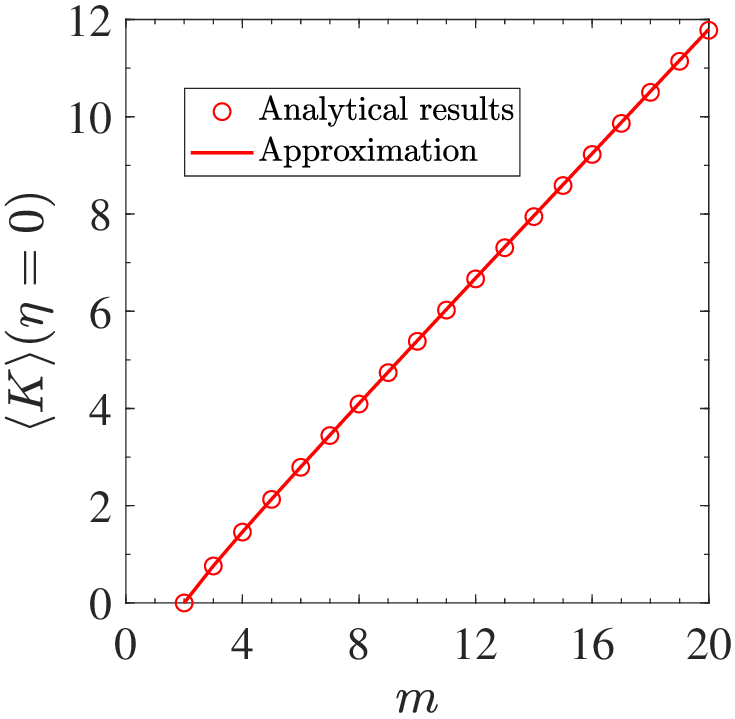} \hspace{0.4 cm}
\end{center}
\caption{
(Color online)
Approximated results for
the mean degree $\langle K \rangle$ as a function of $m$
(solid line),
for the special case of $\eta=0$, obtained from Eq. (\ref{eq:Kmean_for_eta0}).
These results are found to be in very good agreement with the results (circles)
obtained from an exact numerical 
solution of Eq. (\ref{eq:K_SCE}).
}
\label{fig:10}
\end{figure}

In Fig. \ref{fig:10} we present the mean degree $\langle K \rangle$ as a function of $m$
(solid line)
for the special case of $\eta=0$, obtained from Eq. (\ref{eq:Kmean_for_eta0}).
The results are found to be in very good agreement with the exact results (circles)
obtained from a numerical 
solution of Eq. (\ref{eq:K_SCE}).
It is found that the relative error in Eq. (\ref{eq:Kmean_for_eta0}),
compared to the exact results,
is less than one percent for all values of $m > 2$.

\end{document}